\documentclass[prb,twocolumn,superscriptaddress]{revtex4}
\usepackage{amsmath}
\usepackage{graphicx}
\usepackage{amsfonts}
\usepackage{amssymb}
\usepackage{xcolor}
\usepackage{multirow}

\begin{document}

\title{Bulk magnetoelectricity in the hexagonal manganites and ferrites}


\author{Hena Das}
\affiliation{School of Applied and Engineering Physics, Cornell University, Ithaca, NY, USA}
\author{Aleksander L. Wysocki}
\affiliation{School of Applied and Engineering Physics, Cornell University, Ithaca, NY, USA}
\author{Yanan Geng}
\affiliation{Department of Physics and Astronomy, Rutgers University, Piscataway, NJ, USA}
\author{Weida Wu}
\affiliation{Department of Physics and Astronomy, Rutgers University, Piscataway, NJ, USA}
\author{Craig J. Fennie}\email{fennie@cornell.edu}
\affiliation{School of Applied and Engineering Physics, Cornell University, Ithaca, NY, USA}

\date{\today}

\begin{abstract}

Improper ferroelectricity (trimerization) in the  hexagonal manganites RMnO$_3$ leads to  a network of coupled structural and magnetic vortices that induce domain wall magnetoelectricity  and magnetization ($\textbf{M}$) neither of which, however, occurs in the bulk. Here we combined first-principles calculations, group-theoretic techniques, and microscopic spin models  to show how the  trimerization not only induces a polarization ($\textbf{P}$) but also a bulk $\textbf{M}$ and bulk magnetoelectric (ME) effect. This results in the existence of a bulk linear ME vortex structure or a bulk ME coupling such that if $\textbf{P}$ reverses so does $\textbf{M}$. To measure the predicted ME vortex, we suggest  RMnO$_3$ under large magnetic field. We suggest a  family of  materials, the hexagonal  RFeO$_3$ ferrites, also display the predicted phenomena in their ground state.

%

\end{abstract}

\maketitle

\begin{figure*}[t]
\begin{center}
\includegraphics[scale=0.7]{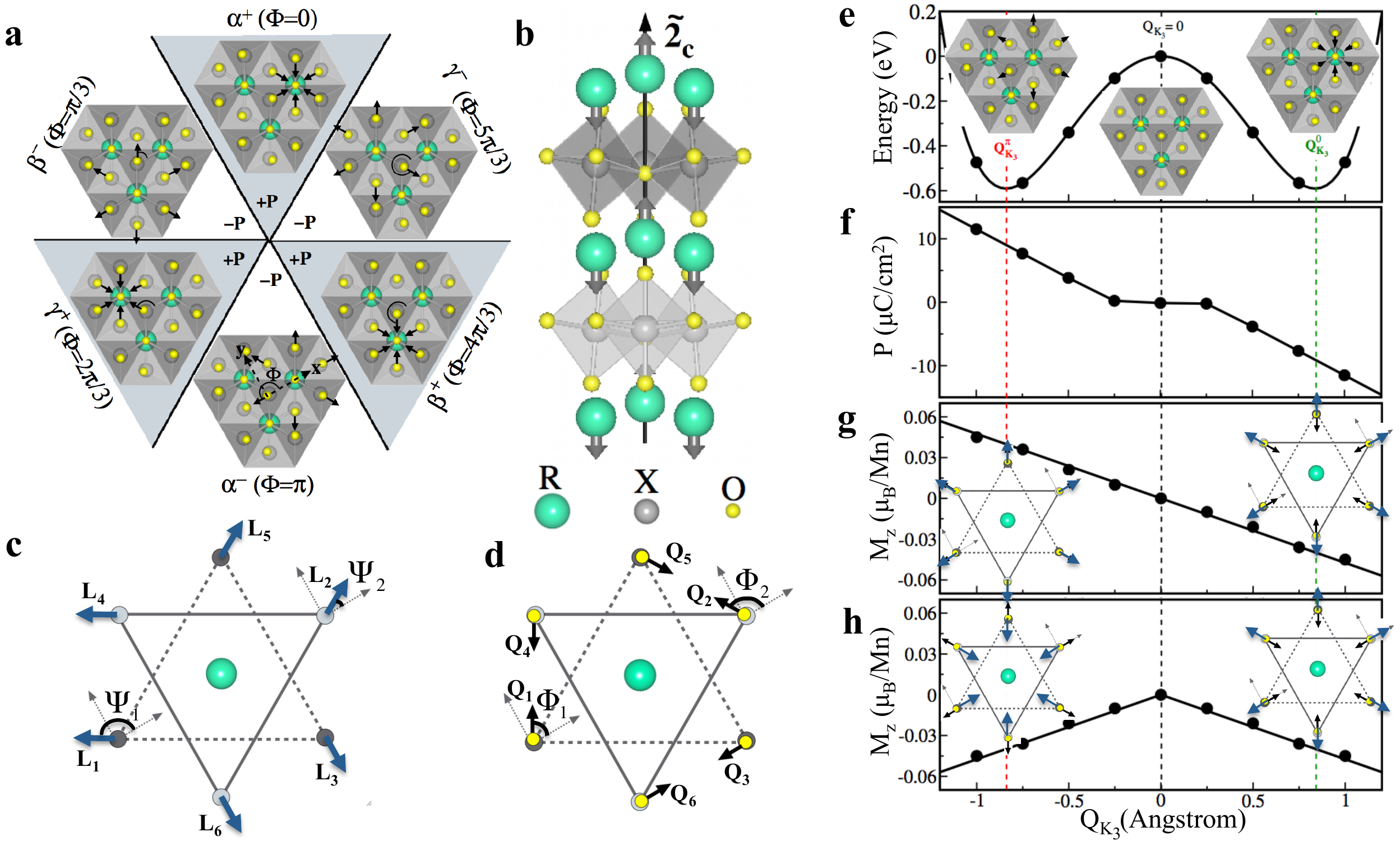}
\end{center}
\caption{\textbf{First Principles: Improper structural ferroelectricity inducing weak-ferromagnetism, \textbf{M}$_z$, in hexagonal structure from first principles.}
Results are shown for ErMnO$_3$, however, they are true for all hexa manganites and RFeO$_3$ ferrites in the A$_2$ phase that we have considered
\textbf{a}, Six  structural  domains of the  primary ``trimer'' $\textbf{Q}_{K_3}^{\Phi}$   distortion  and the secondary polarization \textbf{P}. Here $\alpha^+\Longleftrightarrow\Phi=0$  and a counterclockwise rotation corresponds to domains differing by $\Phi = +\pi /3$. Arrows indicate the direction of the trimer distortions, only distortions around the $\tilde{2}_c$ axis are shown for clarity.
\textbf{b}, crystal structure of the ferroelectric phase ($Q_{K_3}^0$ ($\alpha^+$) domain).
\textbf{c}, The spin angles $\Psi_1$ and $\Psi_2$ describing all possible 120$^\circ$ non-collinear antiferromagnetic (AFM) spin configurations (see Supplemental). $\textbf{L}_i$ ($i\in$ 1 to 6) denotes transverse component of the $i^{th}$ spin. We define $\textbf{L}_1 = \cos\Psi_1 \hat{x}+\sin\Psi_1 \hat{y}$, $\textbf{L}_2 = \cos\Psi_2 \hat{x}+\sin\Psi_2 \hat{y}$ and remaining four vectors can be created by the crystal symmetry operations.
\textbf{d}, Represents directions of local trimer distortions. The local trimer distortion at site 1 ($\textbf{Q}_1$) and 2 ($\textbf{Q}_2$) are defined as, $\textbf{Q}_1 = \cos\Phi_1 \hat{x}+\sin\Phi_1 \hat{y}$ and $\textbf{Q}_2 = \cos\Phi_2 \hat{x}+\sin\Phi_2 \hat{y}$, respectively. Crystal symmetry implies $\Phi_1 \equiv \Phi$ and $\Phi_2=\pi - \Phi_1$.
\textbf{e}, Energy, \textbf{f},  \textbf{P},   \textbf{g} and \textbf{h},  \textbf{M}$_z$ as a function of $Q_{K_3}$, allowing only $\Phi=0$ or $\pi$ trimer domains. Notice that  \textbf{P} =0 when $\textbf{Q}_{K_3}^{\Phi}$ = 0, suggesting that a proper ferroelectric mechanism in not likely. Note that for higher magnitude of $\textbf{Q}_{K_3}$, $P \propto Q_{K_3}$ and $M_z \propto Q_{K_3}$, thus provide linear ME coupling.
}
\label{fig1}
\end{figure*}

Two themes at the forefront of materials physics  are the cross-coupling of distinct types of ferroic order~\cite{bousquet08, tokunaga09, lee10,tokunaga12} and topological defects in systems with spontaneous  broken symmetry~\cite{mermin79,balke12,meier12,tagantsev89}. Common to both are a plethora of novel  phenomenon to understand, and new properties and functionalities to  exploit  for novel applications.  Multiferroics~\cite{ramesh07,cheong07} are an ideal platform to realize both themes in a single material. In this regard, an exciting  development is the  discovery of a topologically protected vortex-domain structure in one of the most extensively studied class of multiferroics, the hexagonal (hexa) rare-earth manganites. Here, antiphase structural (`trimer') domains  are clamped to ferroelectric domain walls (and vice versa)~\cite{choi10,chae10,mostovoy10,kumagai12} forming a `clover-leaf'  pattern, Fig.~\ref{fig1}a. These trimer domains have a particular phase relationship that result in the appearance of structural vortices, which in turn induce magnetic vortices~\cite{geng12,artyukhin12},  strongly coupled antiferromagnetism and the polarization at the domain wall. This domain wall magnetoelectric phenomenon produces a  magnetization localized at the wall~\cite{geng12,artyukhin12}.

The key to realizing these unusual effects is the improper nature of ferroelectricity. Here the polarization ($\textbf{P}$) which is stable in the paraelectric (PE) P6$_3$/mmc structure, is induced by a zone-tripling structural distortion, $\textbf{Q}_{K_3}^{\Phi}$~\cite{aken04,fennie05,artyukhin12}. The latter,  referred to as the trimer distortion,  is  associated with a 2-up/1-down buckling of the R-planes and  tilting of the MnO$_5$ bipyramids, Fig.~\ref{fig1}b. It is nonlinearly coupled to the  polarization,
\begin{equation}
\mathcal{F}_{\rm trimer} \sim {P_z} {Q}_{K_3}^3 { \rm cos}(3 \Phi)
\label{pk3}
\end{equation}
 the form of which implies that a nonzero trimer distortion induces a nonzero $\textbf{P}$. There are three distinct $\Phi$ domains ($\alpha$, $\beta$, and $\gamma$) corresponding to one of the 3 permutations of 2-up/1-down. Also there are two  tilting directions, either towards (+) or away from ($-$) the  $\tilde{2}_c$ axis, i.e., 1-up/2-down or 2-up/1-down, respectively. This results in six  P6$_3$cm structural domains.  A  consequence of the improper origin of ferroelectricity is that the sign of $\textbf{P}$ depends on the direction of  $\textbf{Q}_{K_3}^{\Phi}$. This simple fact leads to the nontrivial domain structure of the hexa manganites, Fig~\ref{fig1}a~\cite{choi10,artyukhin12,mostovoy10, chae10}.

Our focus here is on elucidating a remarkable  interplay of this trimerization, magnetism, and polarization in the hexa manganite structure.
We show from first principles that the trimer structural distortion not only induces a $\textbf{P}$, but can also induce  both a {\it bulk}  magnetization, $\textbf{M}$,  and a {\it bulk} linear ME effect.
We make this clear by connecting an exact microscopic theory of spin-lattice coupling to a simple phenomenological theory.
This  not only brings additional insight to known experiments, but  leads us to discover entirely new bulk phenomena, not previously seen in a multiferroic such as 1) the existence of a linear magnetoelectric (ME) vortex structure and 2) a bulk  coupling of ferroelectric (FE) and ferromagnetic domains such that if $\textbf{P}$ reverses 180$^{\circ}$ so does $\textbf{M}$.

We show that the former is widely accessible in most hexa RMnO$_3$ manganites under large magnetic fields, while the latter is realizable in the ground state of a  new family of  materials, the hexa RFeO$_3$ ferrites.
Recently thin films of RFeO$_3$ have been epitaxially stabilized in the hexa P6$_3$cm structure~\cite{bossak04,magome10, wang13}. By performing a detail comparison of the ferrites to the manganites we  explain why the ground state of all ferrites display an intrinsic, bulk $\textbf{M}$ and  {\it bulk} linear ME effect.

\section{\bf First-principles calculations on switching $\Phi$ by 180$^{\circ}$}
Geometric frustration of the strongly antiferromagnetic (AFM) nearest neighbor Mn (or Fe) spins leads to a planar 120$^{\circ}$  non-collinear order which can be described by two free parameters, $\Psi_1$ and $\Psi_2$, as shown in  Fig.~\ref{fig1}c. Four principle configurations, denoted  A$_1$ ($-\pi/2,\pi/2$), A$_2$ ($\pi,0$), B$_1$ ($0,0$) and B$_2$ ($\pi/2,\pi/2$),  have been previously defined, among which it is well-known that only the A$_2$ (magnetic space group P6$_3$c$^{\prime}$m$^{\prime}$) and   intermediate spin configurations that contain a component of A$_2$ allow a net $\textbf{M}$ along the $z$ axis.
What hasn't been appreciated in the past is that  this known symmetry-allowed $\textbf{M}$ is in fact induced by the trimer distortion, the consequences of which are quite profound.

We now begin to make this clear by performing first-principles calculations for a specific example,  ErMnO$_3$ in the A$_2$ phase (which is the spin configuration realized experimentally under an applied magnetic field~\cite{fiebig03}).
In first-principles calculations of the hexagonal manganite structure it is easy to reverse  the trimer distortion, and hence $\textbf{P}$, via a structural change from a  1-up/2-down buckling and  tilting `in' of the R-planes and MnO$_5$ bipyramids, respectively, to a 2-up/1-down and tilting `out', thus remaining in the same distinct domain, e.g., $\alpha^+\rightarrow \alpha^-$. 

Having performed these calculations we find that the trimer distortion not only induces ferroelectricity,  but  also weak-ferromagetism~\cite{dzyaloshinskii57,moriya60} and the  linear ME effect,  $\alpha_{zz} =\partial M_z/\partial E_z \approx \partial_Q M_z*\partial_Q P_z$ (see Supplemental), as shown in Figures~\ref{fig1}e through \ref{fig1}h. Furthermore, notice that reversal of the trimer distortion reverses either the direction of the $\textbf{M}$, Fig.~\ref{fig1}g, or the sign of the linear ME  tensor, Fig.~\ref{fig1}h,  where both  situations are symmetry equivalent and correspond to whether or not the AFM spin configuration remains fixed, respectively.
 This result is true for all hexa manganites and RFeO$_3$ ferrites in the A$_2$ phase that we have considered and as we prove below is a general property of the A$_2$-phase.

 We pause now to stress the point that real switching will occur via a rotation to a neighboring trimer domain\textcolor{blue}{\cite{artyukhin12}}, e.g., $\alpha^+\rightarrow \beta^-$ or $\gamma^-$. These  first-principles results, however, contain all of the unique ME physics, that is, if the polarization is reversed either the bulk magnetization will reverse or the antiferromagnetic order will change in such a way that the sign of the bulk magnetoelectric tensor changes sign. To understand the consequences of these two choices we next derive a simple phenomenological theory -- valid for  any trimer domain, $\Phi$,  and any spin configuration -- starting from a microscopic model.

\section{\bf Phenomenology theory from microscopic model: Generalizing the first-principles results to switching $\Phi$ by $n{\pi\over 3}$}

 We start by deriving a spin-lattice model from an effective spin Hamiltonian~\cite{solovyev12}
\begin{equation}
H=\sum_{ij}J_{ij}\mathbf{S}_i\cdot \mathbf{S}_j+\sum_{ij}\mathbf{D}_{ij}\cdot \mathbf{S}_j\times \mathbf{S}_j+\sum_i \mathbf{S}_i\cdot \mathbf{\hat{\tau}}_i\cdot \mathbf{S}_i
\label{Ham}
\end{equation}
where the $J_{ij}$'s are the symmetric exchange interactions and  $\mathbf{D}_{ij}$'s are the Dzyaloshinskii-Moriya (DM) antisymmetric exchange vectors, and $\mathbf{\hat \tau}_i$ is the single-ion anisotropy (SIA)  tensor. (Note that our calculations reveal that a dominant DM interaction, $\textbf{D}_{ij}$, exists only between nearest neighbor spins within the triangular planes.)

In  the PE  structure the DM vector has only a $z$ component, which further acts to confine the spins within the $xy$ plane, while the SIA tensor is diagonal. In the FE structure, however, the trimer distortion {\it induces}  a transverse component of the DM vector, $\textbf{D}^{xy}_{ij}$, parallel to the $xy$ plane and off-diagonal components of the SIA tensor. These induced interactions are key and therefore are the focus in the remaining discussion (all other interactions can be safely ignored).

{\bf The effective DM and SIA interactions for a single layer of spins.}
Let us first consider a single layer of spins, denoted as layer {\bf I}. We derive the relationship between the local structural distortions and the induced DM and SIA interactions by considering a single triangle of spins ($\textbf{S}_1$, $\textbf{S}_3$ and $\textbf{S}_5$), in the  $\alpha^+$ and $\alpha^-$ domains, Fig.~S3c (the exact mapping from a single layer of spins to a single triangle is proved in the Supplemental).

Note that the induced DM and SIA interactions cant the spins out of the plane, but all spins in a single plane have to cant in the same direction, we can therefore write $\textbf{S}_i = \textbf{L}_i + \textbf{M}_{\rm I}$, where $\textbf{L}_i\equiv \textbf{S}_i^{xy}$, i.e., the component of the spin lying in the $xy$ plane (Fig.~\ref{fig1}c), while  $\textbf{M}_{\rm I}\equiv \textbf{S}_i^z$ is the net magnetic moment per spin of layer {\bf I}. The total canting energy per spin can be written as
\begin{equation}
E_{\rm I}^{\rm canting}=E_{\rm I}^{\rm DM}+E_{\rm I}^{\rm SIA} = \frac{1}{3}\sum_{i=1,3,5} \mathbf{d}^{\rm eff}_{i}\cdot [ \mathbf{L}_i\times  \mathbf{M}_{\rm I}].
\label{dtotal}
\end{equation}
where $\mathbf{d}^{\rm eff}_{i}=\mathbf{d}_{i}^{\rm DM}+\mathbf{d}_{i}^{\rm SIA}$ is the effective DM-like vector induced by the tilting of the bipyrimid. This effective interaction includes contributions from  both the transverse DM interactions and the off-diagonal elements of SIA tensor (see Supplemental for derivation). Considering that  $\textbf{M}_{\rm I}= M_{\rm I}^z \hat{z}$ and  $M_{\rm I}^z \mathbf{d}^{\rm eff}_i\cdot [ \mathbf{L}_i\times  \hat{z}]  =  M_{\rm I}^z  \mathbf{L}_{i}\cdot\left[{\hat z}\times \mathbf{d}^{\rm eff}_i\right]$, the canting energy per spin can be compactly rewritten as
\begin{equation}
E_{\rm I}^{\rm canting} =  \frac{1}{3}|\mathbf{d}|M_{\rm I}^z \sum_{i=1,3,5} \textbf{L}_i \cdot  \hat{\textbf{Q}}_i =   |\mathbf{d}| M_{\rm I}^z (\textbf{L}_{\rm I}\cdot\hat{\textbf{Q}}_{\rm I})
\label{dot_I}
\end{equation}
where we have defined $\hat{\textbf{Q}}_i$ such that $|\mathbf{d}^{\rm eff}_i| \hat{\textbf{Q}}_i\equiv  \hat{z}\times \mathbf{d}^{\rm eff}_i$, and where by symmetry  $|\mathbf{d}^{\rm eff}_i| = |\mathbf{d}^{\rm eff}| \equiv |\mathbf{d}|$  and $\textbf{L}_1 \cdot  \hat{\textbf{Q}}_1 = \textbf{L}_2 \cdot  \hat{\textbf{Q}}_2 = \textbf{L}_3 \cdot  \hat{\textbf{Q}}_3 \equiv \textbf{L}_{\rm I}\cdot\hat{\textbf{Q}}_{\rm I}$.

It is  interesting that  $\mathbf{\hat{Q}}_i$ is the direction of the in-plane displacement of the apical oxygen that lies directly above  spin $\mathbf{S}_i$.  It is zero in the PE phase and is in opposite directions in the  $\alpha^{\pm}$ domains. It behaves in every aspect as an order parameter that defines the local trimer distortion. In fact, one of the $ \mathbf{\hat{Q}}_i$'s is the atomic distortion that Mostovoy has used to define a particular trimer domain~\cite{artyukhin12}. If we had considered a different domain, e.g., $\beta^+$,  the $\mathbf{\hat{Q}}_i$'s rotate appropriately and  in fact have similar transformational properties as the trimer structural distortions, $ \mathbf{Q}^{\Phi}_{K_3}$. This  suggests (and we prove in the Supplemental) that $\mathbf{\hat{Q}}_i$ is a local  trimer distortion, which induces the local DM-like interaction, $\mathbf{d}_i^{\rm eff}$, and subsequently cants the spins.

It is now clear that 1) if the relative phase between the local AFM spin, $\textbf{L}_{\rm I}$, and the local trimer distortion, $\mathbf{\hat{Q}}_{\rm I}$, changes sign, the net magnetic moment per spin of a layer, $\mathbf{M}_{\rm I}$, reverses, \textit{i.e}, $\mathbf{M}_{\rm I} \propto (\mathbf{L}_{\rm I}\cdot\mathbf{\hat{Q}}_{\rm I}){\hat{z}}$, and 2) canting occurs  only in a FE phase (where $\mathbf{\hat{Q}}_{\rm I}\ne0$) and only when there is a nonzero projection of a spin along the direction of the local trimer distortion. This is why there is no canting for the A$_1$ and B$_2$ spin states (where $\mathbf{L}_{\rm I}\cdot\mathbf{Q}_{\rm I} =0$).

{\bf The real  structure -- the stacking of two layers.}
The real hexa unit cell has two triangular planes,  layer {\bf I} (which includes sites 1, 3, 5)  and  layer {\bf II} (which includes sites 2, 4, 6), stacked along the $z$ axis, as shown in Fig.~\ref{fig1}c and d.   The canting energy per spin is
\begin{equation}
E_{\rm canting}=   |\mathbf{d}| \left[M^z_{\rm I} (\textbf{L}_{\rm I}\cdot\mathbf{\hat{Q}}_{\rm I}) + M^z_{\rm II}(\mathbf{L}_{\rm II}\cdot\mathbf{\hat{Q}}_{\rm II})\right]
\label{DMfinal}
\end{equation}
This can be alternatively written as
\begin{eqnarray}
E_{\rm canting} &=&   |\mathbf{d}|M_z[\textbf{L}_{\rm I}\cdot\mathbf{\hat{Q}}_{\rm I} + \mathbf{L}_{\rm II}\cdot\mathbf{\hat{Q}}_{\rm II}] \nonumber\\
&&+  |\mathbf{d}|L_z[\textbf{L}_{\rm I}\cdot\mathbf{\hat{Q}}_{\rm I} - \mathbf{L}_{\rm II}\cdot\mathbf{\hat{Q}}_{\rm II}]
\label{DMfinal2}
\end{eqnarray}
where $M_z\equiv(M_{\rm I}^z+M_{\rm II}^z)/2$ and $L_z\equiv(M_{\rm I}^z-M_{\rm II}^z)/2$ are the total and stagger magnetic moment per spin respectively.
Although the specific sites we choose to define {\bf I} and  {\bf II} are arbitrary, it is convenient to associate  {\bf I} with site  {\bf 1} and {\bf II} with site  {\bf 2}. Because symmetry implies $\Phi_1=\pi-\Phi_{2}$, where $\Phi_1$  ($\Phi_2$) is the local trimer angle at site {\bf 1} ({\bf 2}), a   single trimer angle, $\Phi   \equiv \Phi_1$, can be defined. This single angle was used to define the trimer domains in Fig.~\ref{fig1}a.

 Note that Eq.~\ref{DMfinal2} is the exact result we derived from Landau theory (see Supplemental) and explains our first-principles calculations displayed in Fig.~\ref{fig1}; in the A$_2$ state  $\textbf{L}_{\rm I}\cdot\mathbf{\hat{Q}}_{\rm I}$ = +$\textbf{L}_{\rm II}\cdot\mathbf{\hat{Q}}_{\rm II}$,  and therefore $M^z_{\rm I} = M^z_{\rm II}$ leading to a net magnetization as we previously showed from first principles.
For completeness note that in the  B$_1$ state,  however, the projection has the opposite sign in adjacent layers, $\textbf{L}_{\rm I}\cdot\mathbf{\hat{Q}}_{\rm I}$ = -$\textbf{L}_{\rm II}\cdot\mathbf{\hat{Q}}_{\rm II}$. The spins in each  plane still cant but since the projection  changes sign in adjacent layers no net magnetization exists, $M^z_{\rm I} =- M^z_{\rm II}$. We call this weak-antiferromagentism, wAFM.

\section{\bf Implications: testable predictions}

Note that  if $\textbf{P}$ switched via rotating $\Phi$ by $\pi/3$, e.g.,    $\alpha^+\rightarrow\beta^-$,  $\textbf{L}$ must rotate by either $|\pi/3|$ ($\textbf{L}\cdot\mathbf{\hat{Q}}_{\alpha^+} = -1 \rightarrow\textbf{L}\cdot\mathbf{\hat{Q}}_{\beta^-} =-1$) or $|2\pi/3|$  ($\textbf{L}\cdot\mathbf{\hat{Q}}_{\alpha^+} = -1 \rightarrow\textbf{L}\cdot\mathbf{\hat{Q}}_{\beta^-} = +1$).
s

\begin{figure}[t]
\begin{center}
\includegraphics[scale=0.35]{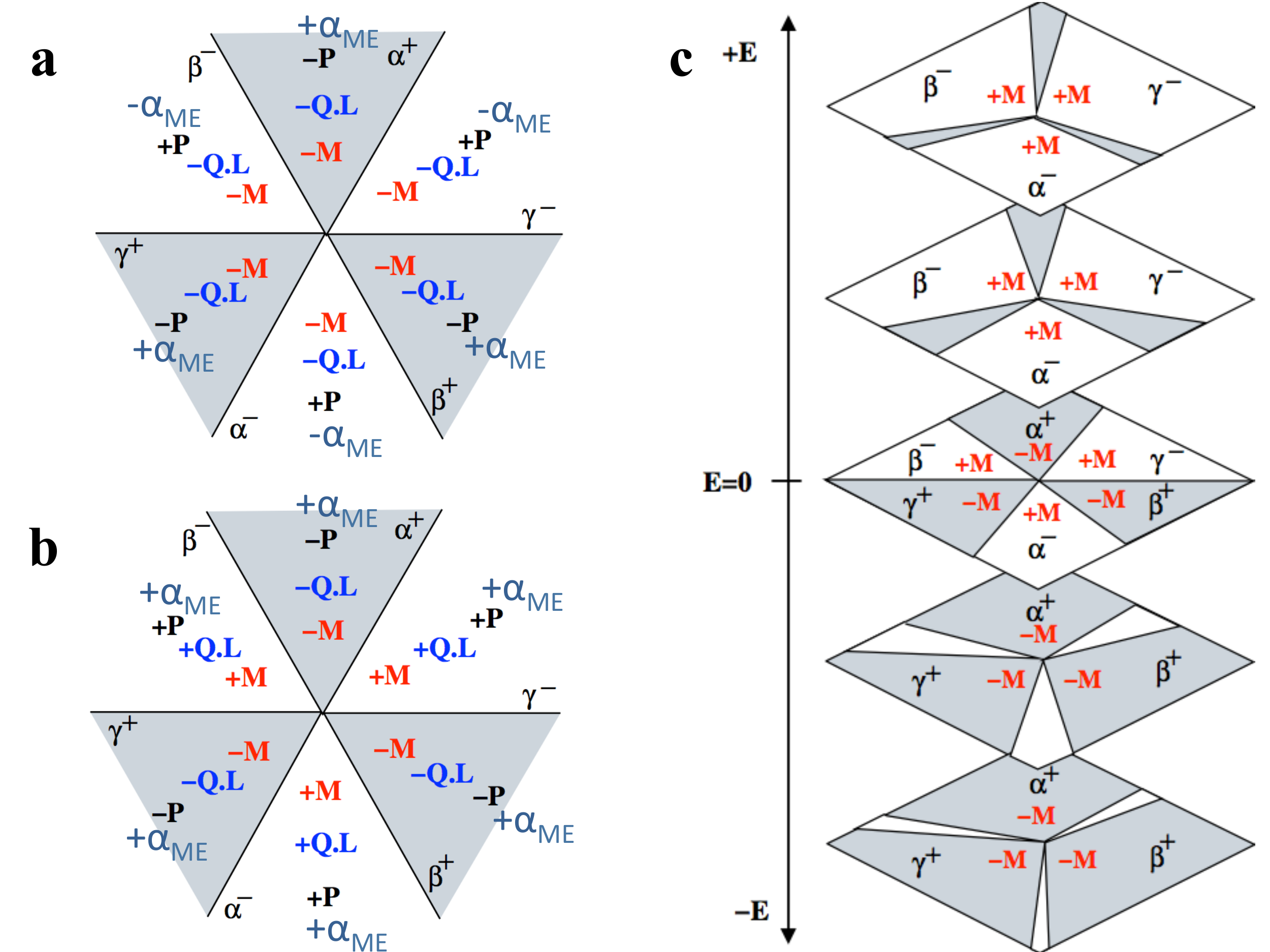}
\end{center}
\caption{\textbf{Predicted clamping of trimer and magnetic domains and electric field response.}  At the lowest energy trimer domain wall $\Phi$ changes by $\pi/3$, which will rotate the AFM spin vectors, $\textbf{L}_1$ and $\textbf{L}_2$, by either  \textbf{a},  $\pi/3$ and $-\pi/3$ or \textbf{b},  $-2\pi/3$ and $2\pi/3$. In the former case  {\bf M}$_z$ remains constant while in the latter  {\bf M}$_z$ reverses at each trimer domain wall, i.e., ferroelectric, ferromagnetic, and trimer domains are mutually clamped. \textbf{c}, The schematic diagram displaying the electric-field switching of  ferromagnetic domains (note, figure inspired by Ref.\cite{choi10}). }
\label{fig2}
\end{figure}

{\it Prediction 1--} In the former case the expected domain configurations, Fig.~\ref{fig2}a, are such that at trimer domain walls differing by $\Delta\Phi=\pi/3$ the AFM  spins rotate $\Delta \Psi_{1,2} = |\pi/3|$. In this case the direction of the magnetization remains the same across the domain wall, i.e., although $\textbf{P}$  switches,  $\textbf{M}$  is not reversed similar to that shown in Fig~\ref{fig1}h. Still, there exist a bulk   linear ME  effect
\begin{equation}
\alpha_{zz} \propto{\rm cos}(3\Phi)\, ( \textbf{L}_{\rm I}\cdot\mathbf{\hat{Q}}_{\rm I} + \textbf{L}_{\rm II}\cdot\mathbf{\hat{Q}}_{\rm II} )
\label{eqME}
\end{equation}
(see Supplemental), which in this case leads to a presence of a remarkable linear ME vortex structure  as the projections, $\textbf{L}_{\rm i}\cdot\mathbf{\hat{Q}}_{\rm i}$,  are equal in all domains, and therefore $\alpha_{zz}$ is of opposite sign in neighboring trimer domains.

{\it Prediction 2--} Note that this seemingly lower energy AFM switching pathway results in a homogeneous magnetization across the entire material. In zero field, this has to be unstable towards the formation of ferromagnetic domains. But what kind of domains? 
They can occur within the bulk of a trimer domain, i.e., a free AFM domain. There is, however, an energy cost to form this domain wall.
An alternative path to minimize the total energy of the system is considered in Fig.~\ref{fig2}b, where at trimer domain walls that differ by $\Delta\Phi=\pi/3$ the AFM  spins now rotate $\Delta \Psi_{1,2} = |2\pi/3|$.  In this case, the magnetization direction reverses with the polarization similar to that shown in Fig~\ref{fig1}g. Therefore, even though this  domain configuration at first appears less likely than Fig~\ref{fig2}a, it provides an avenue for the system to minimize the magnetostatic energy without having to introduce free domains.
Additionally, in Fig.~\ref{fig2}c we sketch the expected response of the domains to electric-field poling.  In this process the positive electric field, $\mathbf{E}$, e.g., chooses the ($+\mathbf{P}$,$+\mathbf{M}$) state and therefore reversing of the direction of electric field  will not only switches the direction of polarization, but also  reverses the direction of magnetization.

\section{\bf Discussion: possible realizations of predictions}

{\bf Realization 1: The hexagonal Manganites }.
%
In the hexagonal manganites, e.g., ErMnO$_3$, an A$_2$ phase appears under the application of external magnetic field~\cite{fiebig03}. Here, as the magnetic field is swept from zero to a large (for example) positive value, a ME vortex structure is expected to appear. Additionally, as the magnetic field is scanned to negative values the sign of the ME vortex structure should switch. This is precisely what our preliminary imaging of the ME vortex structure shows using a new technique called Magnetoelectric Force Microscopy, as shown in the supplemental.

{\bf Realization 2:  The A$_2$ ground state in hexa-RFeO$_3$}.
%
Are there materials in which this physics is realized in the ground state?
Recently thin films of RFeO$_3$ have been epitaxially stabilized in the hexa P6$_3$cm structure~\cite{bossak04,magome10}. These hexa ferrites  exhibit ferroelectricity above room temperature, but with conflicting results as to its origin~\cite{jeong12a,jeong12b}.  Additionally there is  evidence of a  magnetic transition around $\sim$ 100K, at which $\textbf{M}$ becomes nonzero~\cite{jeong12a,jeong12b,akbashev11,wang13}, however, the significance of this or even if it is an intrinsic or bulk effect is not previously  known.

Our  calculations suggest strongly that ferroelectricity in the hexa ferrites is of the improper structural type where the trimer distortion induces $\textbf{P}$ (see Supplemental for full discussion),  and therefore a similar topological domain structure should exist as in the manganites.  The difference in electronic structure between manganites and ferrites, however, requires Fe spins of any hexa ferrite to order in the A$_2$ spin configuration in ground state. Additionally, the much stronger exchange interactions leads to the possibility of  spin ordering above room temperature, as recently suggested by the experiments of Ref.~\onlinecite{wang13}.)
Therefore, both scenarios displayed in  Fig.~\ref{fig2} are possible. We now discuss our first-principles calculations indicating that the ground state of any hexagonal ferrite will indeed have the A$_2$ magnetic configuration.

\begin{figure}
\begin{center}
\includegraphics[scale=0.35]{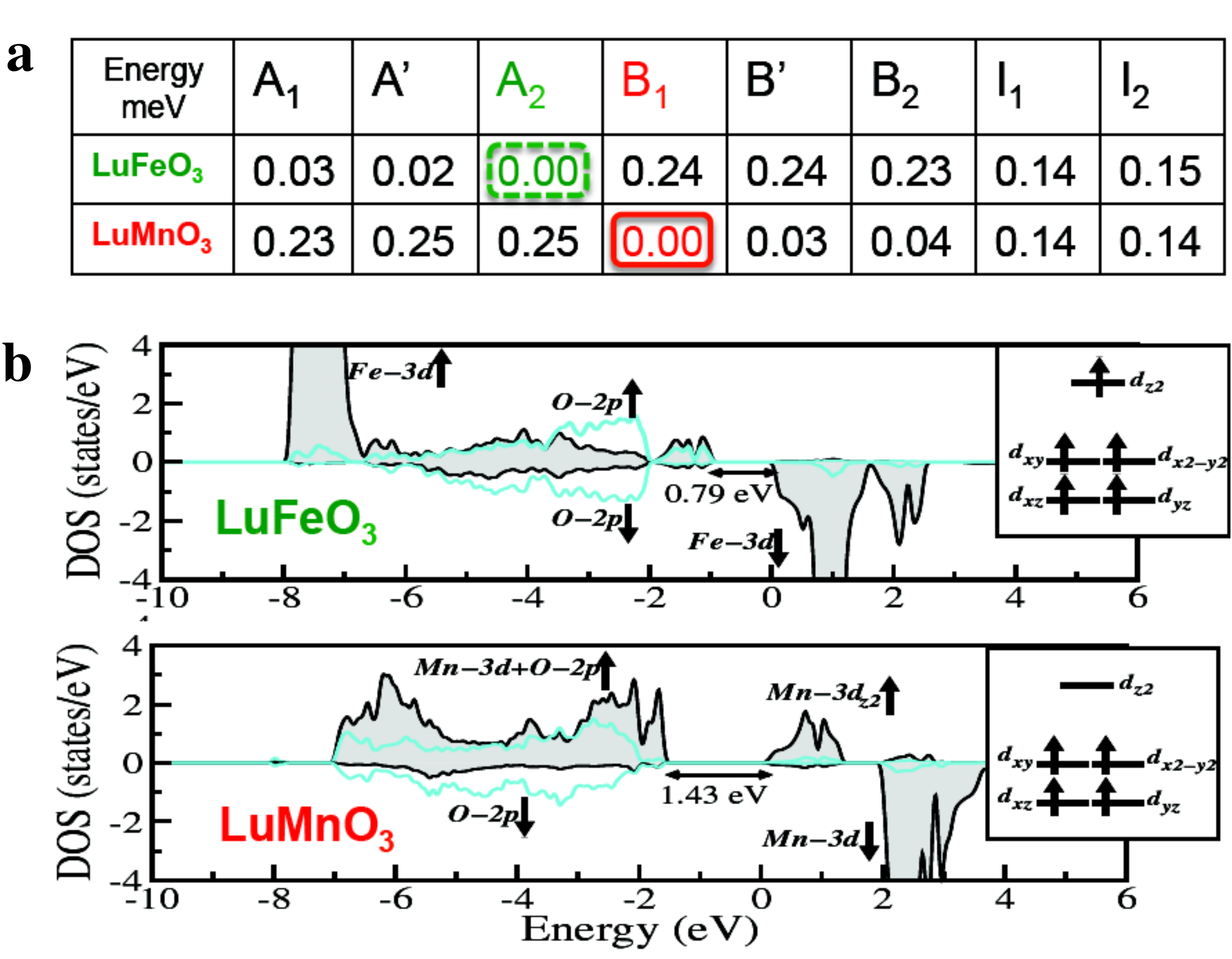}
\end{center}
\caption{\textbf{The noncollinear magnetic and electronic structure from first principles.}
First-principles calculations of \textbf{a}, Noncollinear magnetic energies and \textbf{b}, density of states (collinear). Note that the difference in orbital occupancy is  the fundamental reason why ferrites, top,  and manganites, bottom, have different magnetic ground states and subsequently the key to the predicted  bulk coupling of polarization and magnetism in ferrites. Insets: crystal field splitting and occupancy of the TM majority $3d$ channel.
}
\label{fig3}
\end{figure}

{\it Magnetic structure.} In addition to the  principle magnetic configurations we also considered the four known intermediate magnetic structures (see Fig.S7). The results of our total energy calculations for LuFeO$_3$, LFO,  and LuMnO$_3$, LMO, are presented in Fig.~\ref{fig3}a (for clarity we limit our discussion to these two compounds). In agreement with non-linear optical measurements~\cite{fiebig00}, we find that LMO stabilizes in the wAFM B$_1$ state.  In contrast,  LFO stabilizes in the wFM A$_2$ state giving rise to a  net canted spin moment $M_z=$ 0.02$\mu_B$/Fe  along  the $z$ axis. Note, however, that the A$_1$ state, where the net magnetic moment is equal to zero in each layer by symmetry, is also close in energy.
%
%
%
%

{\it Electronic structure.} In the PE  phase the crystal field at the TM site has a D$_{3h}$ trigonal point symmetry, which splits atomic 3$d$ levels into three sets of states as shown in the insets of Fig.\ref{fig3}b.
The density of states (DOS) plots calculated for LFO and LMO  in the FE phase are shown in Fig.~\ref{fig3}b. LFO is a charge-transfer insulator with the conduction band formed by minority Fe 3$d$ states and the valence band composed of O 2$p$ states, below which are the filled majority Fe 3$d$ bands. In the case of LMO majority 3$d$ bands are partially filled with $d^4$ electronic configure, while minority 3$d$ levels are completely empty. The importance of these differences will be made clear when discussing the magnetic interactions.

%

{\it Symmetric exchange.}  Although it is the DM interactions and SIA that drives spin canting, it is the symmetric exchange that determines the magnetic configuration type.
There are two  important symmetric exchange interactions. The first is a strong, AFM superexchange interaction between in-plane  nn spins, $J_{nn}$. Its magnitude  is much larger for LFO compared with LMO, suggesting a substantially larger magnetic ordering temperature for ferrites (within mean field theory the calculated Currie-Weiss temperature for LFO in the FE phase is $\Theta_{CW} =1525$ K, while in LMO  $\Theta_{CW} =274$ K~\cite{comment_U}).

The second is a weak, super-superexchange  interaction, $J_c$,  which couples consecutive spin planes via a TM-O-Lu-O-TM exchange pathway (see Fig.S8a).  In the PE  structure, a spin in one layer is connected to three spins in a consecutive layer. Each of these degenerate spin-spin interactions has two equivalent exchange pathways. We find that the interlayer exchange is AFM for LFO but FM for LMO. Although the strength of this interaction is relatively weak, this sign difference turns out to be key.

In the PE structure  symmetry implies that the relative orientation of the spins in  consecutive spin planes is arbitrary. The trimer distortion, however, splits the three degenerate interactions into:  a single $J_c^{11}$ interaction, mediated by two equivalent TM-O-Lu1-O-TM exchange pathways, and two $J_c^{12}$ interactions, where each interaction is mediated by a  TM-O-Lu1-O-TM and a  TM-O-Lu2-O-TM exchange pathway (see Fig.S8b). This  remarkably introduces an extra contribution to the energy
\begin{equation}
E_{\rm inter}^{\rm SE}=2\Delta J_{c}\cos(\Psi_1-\Psi_2)
\end{equation}
 where $\Delta J_c=J_c^{11}-J_c^{12}$, the sign of which is key in determining the spin configuration type:  A-type ($\Psi_2$=$\Psi_1$+$\pi$) for $\Delta J_c>0$ or B-type ($\Psi_2$=$\Psi_1$) for $\Delta J_c<0$.

A simple structural analysis shows that the super-super exchange mediated through the Lu$_2$ ion is always weaker than that mediated through the Lu$_1$ ion, and indeed our calculations show that the magnitude of $J_c^{11}$ is always larger than $J^{12}_c$ (see Supplemental). We therefore see that the choice between A-type and B-type in the FE structure is in fact determined by the sign of $J_c$ in the PE structure. This is important. In ferrites,   the AFM nature of the interlayer exchange is uniquely determined by the orbital occupancy, it is always AFM and therefore ferrites will always prefer $A$-type magnetic configurations and the wFM ground state. (Although the interlayer exchange in LMO is FM, which explains why it prefers $B$-type magnetic configurations, it is not universally so; a  discussion is given in the Supplement).


\section{Summary}
%
In this paper we have discussed an intriguing consequence of improper ferroelectricity in the hexagonal manganite-like systems. We have shown for the first time that a non-polar  trimer structural distortion not only induces an electrical polarization but also induces bulk weak-ferromagnetism and a bulk linear magnetoelectric vortex structure.  It is a universal feature of A$_2$-type hexa systems in which the trimer distortion mediates an intrinsic bulk trilinear-coupling of the polarization, magnetization, and antiferromagnetic order.

Note that it was recently inferred from neutron diffraction~\cite{wang13}  that LFO orders  above room temperature (high for a frustrated magnet)  in an AFM state with $\textbf{M}=0$, and at a lower temperature undergoes a reorientation transition to the A$^{\prime}$ phase inducing a $\textbf{M}\ne0$. As shown in Fig.~\ref{fig3}a, the A$_1$  ($\textbf{M}=0$ by symmetry) and A$^{\prime}$ (finite $\textbf{M}$ allowed) states, lie energetically very close to ground state in LFO, which support such a picture.

There is, however, an intriguing alternative scenario involving a crossover from a state in which several magnetic order types are degenerate to the A$_2$ spin configuration ground state, driven by the trimer distortion. Note that the symmetry of the PE structure not only  implies that the A and B spin configurations are degenerate, but in fact that all of the principle spin configurations are degenerate as there can be no in plane anisotropy. The trimer distortion lifts the degeneracy between the A and B spin configurations as
\begin{equation}
\Delta J_c \propto Q_{K_3}^2
\end{equation}
 while the in plane anisotropy due to the effective DM-like interaction (the trimer distortion induced in plane anisotropy of the SIA tensor is negligible) always favors phases with canted spins and is lifted as 
\begin{equation}
\mathbf{d}^{\rm eff}\propto Q_{K_3}
\end{equation}
 as we show in the Supplemental.
In this picture, as temperature is lowered and the trimer distortion  increases in magnitude, there is a smooth crossover to the A$_2$ state.

\section{ \bf Method} The first principles calculations were performed using the DFT+U method~\cite{anisimov97} with the PBE form of exchange correlation functional~\cite{PBE}. We considered Lu $4f$ states in the core and for TM $3d$ states we chose $U=4.5$ eV and $J_H=0.95$ eV. Structural relaxations, frozen phonon and electric polarization calculations  were performed without the spin-orbit coupling (SOC) using the projected augmented plane-wave basis based method as implemented in the VASP~\cite{VASP1,VASP2}.  We used a 4$\times$4$\times$2 k-point mesh and a kinetic energy cut-off of 500 eV. The Hellman-Feynman forces were converged to 0.001 eV/\AA. The electronic and magnetic properties were  studied in the presence of SOC. We additionally cross-validated the electronic and magnetic properties using the Full-potential Linear Augmented Plane Wave (FLAPW) method as implemented in WIEN2K code~\cite{wien2k}.\\

\section{\textbf{Supplementary materials}} 
\begin{figure*}[t]
\begin{center}
\includegraphics[scale=0.75]{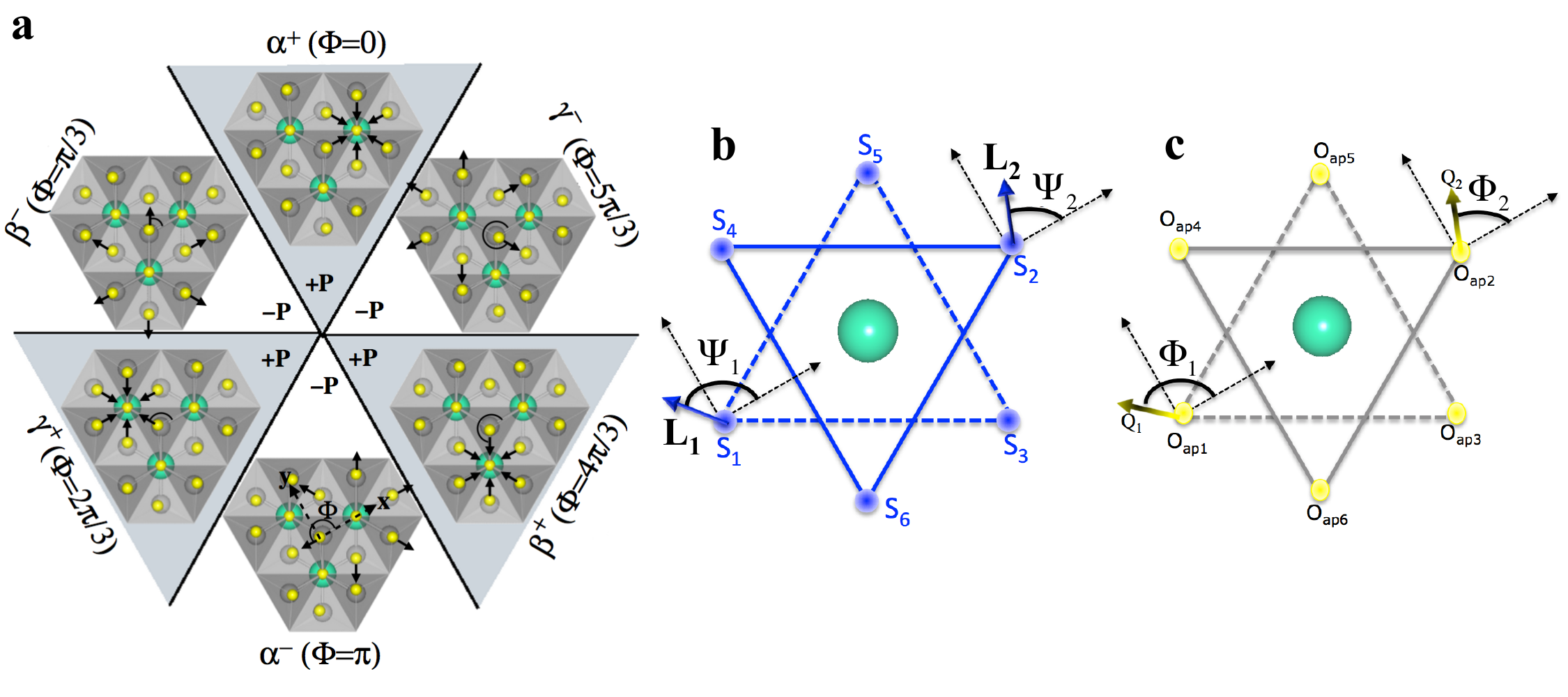}
\end{center}
\caption{\textbf{Magnetic and trimer order parameters.}
\textbf{a} "Clover-leaf" structural domain pattern, where antiphase structural domains are clamped to ferroelectric domain walls. \textbf{b} The spin angles $\Psi_1$ ($\equiv \Psi_{\rm I}$) and $\Psi_2$ ($\equiv \Psi_{\rm II}$), describing the in-plane spin directions for two reference TM spins S$_1$ and S$_2$ respectively.
\textbf{c} Phase angles $\Phi_1$ ($\equiv \Phi_{\rm I}$) and $\Phi_2$ ($\equiv \Phi_{\rm II}$) that describe the local trimer distortion for layer I and II respectively, where $\Phi_1=\Phi$ and $\Phi_2=\pi-\Phi_1$.}
\label{Fig1_supp}
\end{figure*}

%
\subsection{Landau Theory}
\label{landau}
%

In this section we use a phenomenological Landau theory to show the existence of a trilinear coupling between the antiferromagnetic order, the trimer distortion and the magnetization.

The unit cell contains two triangular layers of transition metal (TM) ions. Within the layer $\alpha$ ($\alpha=$I, II) a local magnetic structure is a combination of 120$^{\circ}$ antiferromagnetic order in the $xy$ plane and a magnetization along the $z$ axis, $M_{\alpha}^z$. The former can be represented by a complex order parameter
\begin{equation}
L_{\alpha}=L_{xy}e^{i\Psi_{\alpha}}
\end{equation}
where the angles $\Psi_{\rm I}$ and $\Psi_{\rm II}$ describe in-plane spin directions of two reference TM ions, as we consider $S_1$ and $S_2$ respectively, from adjacent layers connected by the $\tilde2_c$ axis that remains in the FE phase, see Fig.~\ref{Fig1_supp}b. Therefore $\Psi_{\rm I}\equiv \Psi_1$ and $\Psi_{\rm II}\equiv \Psi_2$. Note that because of symmetry we only need to consider one of the three spins in each layer.

The layer magnetizations can be alternatively represented by the $z$ components of the net magnetization ($M_z\equiv (M_{\rm I}^z+M_{\rm II}^z)/2$) and staggered magnetization ($L_z\equiv (M_{\rm I}^z-M_{\rm II}^z)/2$).

The trimer distortion corresponds to the condensation of the zone-boundary $K_3$ mode. While the small representation of $K_3$ is one-dimensional, the star contains two wavevectors ($\mathbf{k}$ and $-\mathbf{k}$) and therefore the trimer distortion can be described by a complex order parameter
\begin{equation}
Q_{\rm I}=Q_{K_3}e^{i\Phi_{1}}
\end{equation}
It turns out that $\Phi_{1}\equiv\Phi$ transforms as the phase angle that describes the in-plane displacement of some reference apical oxygen in the FE phase which we choose to lie directly above the reference TM ion used for defining $\Psi_1$, , see Fig.~\ref{Fig1_supp}b and c. While the order parameter $Q_{\rm I}$ fully describes the trimer distortion, it is convenient to introduce additional trimer order parameter
\begin{equation}
Q_{\rm II}=Q_{K_3}e^{i\Phi_{2}}
\end{equation}
with $\Phi_2=\pi-\Phi_1$. As shown in Fig.~\ref{Fig1_supp}c, $\Phi_2$ describes the in-plane displacement of the apical oxygen lying directly above the reference TM ion used for defining $\Psi_2$. The introduction of the second trimer order parameter allows us to represent the structural distortion in an analogous way as the magnetic ordering which will lead to a particularly transparent form of coupling between structure and magnetism.

The character table below shows the transformation properties of the $L_{\alpha}$ and $Q_{\alpha}$ order parameters as well as their complex conjugates with respect to symmetry operations of the P6$_3$/mmc1' reference structure (included are only the symmetry elements that are broken by the magnetic and/or the trimer orderings). In addition, the transformation properties of the following combinations of these order parameters are shown
\begin{equation}
X_{\eta}=Q_{\rm I}L^{*}_{\rm I} +\eta Q_{\rm II} L^{*}_{\rm II}
\end{equation}
where $\eta=\pm$. Note that $X_{\eta}$ are the only bilinear combinations of antiferromagnetic and trimer order parameters that are invariant under translation. The transformation properties of $M_z$, $L_z$, and the $z$ components of the electric polarization ($P_z$) are also shown.

\begin{table}[h]
\caption{The character table showing the transformation properties of $L_{\alpha}$ and  $Q_{\alpha}$ order parameters as well as their complex conjugates with respect to symmetry operations of the P6$_3$/mmc1' reference structure. In addition the transformation properties of $X_{\eta}$, $P_z$, $M_z$, and $L_z$ are shown. We defined $\eta=\pm$ and $\phi=e^{-i2\pi/3}$}.
\begin{center}
\begin{tabular}{|c|c|c|c|c|c|}
\hline
 &$S_x$  &$\tilde2_c$ &$m_{xy}$ &$I$ &$R$ \\
\hline
$P_z$                   &$+P_z$                              &$+P_z$                  &$+P_z$                    &$-P_z$                      &$+P_z$ \\
\hline
$Q_{\rm I}$     &$\phi Q_{\rm I}$           &$-Q_{\rm II}$    &$+Q^*_{\rm I}$  &$+Q_{\rm II}$    &$+Q_{\rm I}$ \\
$Q^*_1$ &$\phi^{*}Q^*_{\rm I}$ &$-Q^*_{\rm II}$ &$+Q_{\rm I}$     &$+Q^*_{\rm II}$ &$+Q^*_{\rm I}$ \\
$Q_{\rm II}$     &$\phi^{*}Q_{\rm II}$    &$-Q_{\rm I}$     &$+Q^*_{\rm II}$ &$+Q_{\rm I}$     &$+Q_{\rm II}$ \\
$Q^*_{\rm II}$ &$\phi Q^*_{\rm II}$       &$-Q^*_{\rm I}$ &$+Q_{\rm II}$     &$+Q^*_{\rm I}$  &$+Q^*_{\rm II}$ \\
\hline
$L_{\rm I}$     &$\phi L_{\rm I}$            &$-L_{\rm II}$     &$-L^*_{\rm I}$   &$+L_{\rm II}$      &$-L_{\rm I}$  \\
$L^*_{\rm I}$ &$\phi^{*}L^*_{\rm I}$  &$-L^*_{\rm II}$  &$-L_{\rm I}$      &$+L^*_{\rm II}$   &$-L^*_{\rm I}$ \\
$L_{\rm II}$     &$\phi^{*}L_{\rm II}$     &$-L_{\rm I}$      &$-L^*_{\rm II}$   &$+L_{\rm I}$      &$-L_{\rm II}$ \\
$L^*_{\rm II}$ &$\phi L^*_{\rm II}$        &$-L^*_{\rm I}$  &$-L_{\rm II}$       &$+L^*_{\rm I}$   &$-L^*_{\rm II}$ \\
\hline
$X_{\eta}$       &$+X_{\eta}$                   &$\eta X_{\eta}$ &$-X^{*}_{\eta}$          &$\eta X_{\eta}$    &$-X_{\eta}$ \\
\hline
$M_z$                  &$+M_z$                              &$+M_z$                 &$-M_z$                    &$+M_z$                     &$-M_z$ \\
$L_z$               &$+L_z$                          &$-L_z$               &$-L_z$                 &$-L_z $                 &$-L_z$  \\
\hline
\end{tabular}
\end{center}
\label{t1}
\end{table}

From the above table it is clear that the following two free energy invariants are allowed:
\begin{equation}
F^{M_z}_{\text{tri}}\sim \Re[X_+]M_z
\end{equation}
\begin{equation}
F^{L_z}_{\text{tri}}\sim \Re[X_{-}]L_z
\end{equation}
where $\Re$ denotes a real part. These invariants can be alternatively written as
\begin{eqnarray}
F^{M_z}_{\text{tri}}&\propto& (\mathbf{Q}_{\rm I}\cdot \mathbf{L}_{\rm I} + \mathbf{Q}_{\rm II}\cdot\mathbf{L}_{\rm II}) M_z
\label{m}\\
F^{L_z}_{\text{tri}}&\propto& (\mathbf{Q}_{\rm I}\cdot \mathbf{L}_{\rm I} - \mathbf{Q}_{\rm II}\cdot\mathbf{L}_{\rm II})L_z\label{l}
\end{eqnarray}
where we defined two-dimensional real vectors $\mathbf{L}_{\alpha}=L_{xy}(\cos{\Psi_{\alpha}},\sin{\Psi_{\alpha}})$ and $\mathbf{Q}_{\alpha}=Q_{K_3}(\cos{\Phi_{\alpha}},\sin{\Phi_{\alpha}})$. As a consequence of this trilinear coupling, presence of the trimer distortion and the 120$^{\circ}$ antiferromagnetic order leads to weak (anti)ferromagnetism with (staggered) magnetization given by
\begin{eqnarray}
M_z&\propto& (\mathbf{Q}_{\rm I}\cdot \mathbf{L}_{\rm I} + \mathbf{Q}_{\rm II}\cdot\mathbf{L}_{\rm II})
\label{inv1}\\
L_z&\propto& (\mathbf{Q}_{\rm I}\cdot \mathbf{L}_{\rm I} - \mathbf{Q}_{\rm II}\cdot\mathbf{L}_{\rm II})
\label{inv2}
\end{eqnarray}

Note that due to equivalence of TM ion layers we have $|M_{\rm I}^z|=|M_{\rm II}^z|$ which in turn implies that the proportionality coefficients in Eqs. (\ref{inv1}) and (\ref{inv2}) are equal. Therefore, by adding and subtracting Eqs. (\ref{inv1}) and (\ref{inv2}) we obtain
\begin{equation}
M_{\alpha}^z\propto \mathbf{Q}_{\alpha}\cdot \mathbf{L}_{\alpha}
\label{Mz_Landau}
\end{equation}
where the proportionality coefficient doesn't depend on $\alpha$. The above equation clearly shows that for a given TM ion layer canting appears when a projection of an $xy$-spin along the direction of the local trimer distortion is nonzero which is for any spin configuration except A$_1$ and B$_2$ structures. Further, if the projection have the same sign for adjacent layers (A spin configurations) there is a net magnetization while if for neighboring layers the projections have opposite signs (B spin configurations) we have weak anti ferromagnetism.

In order to make a connection of the phenomenological theory with our microscopic model let's define the vector $\mathbf{d}^{\rm eff}_{\alpha}$ by $\mathbf{Q}_{\alpha}=\hat{\mathbf{z}}\times\mathbf{d}^{\rm eff}_{\alpha}$. Then invariants (\ref{m}) and (\ref{l}) can be written as
\begin{eqnarray}
F^{M_z}_{\text{tri}}&\propto& (\mathbf{d}^{\rm eff}_{\rm I}\cdot(\mathbf{L}_{\rm I}\times\hat{\mathbf{z}})+\mathbf{d}^{\rm eff}_{\rm II}\cdot(\mathbf{L}_{\rm II}\times\hat{\mathbf{z}})) M_z \\
F^{L_z}_{\text{tri}}&\propto& (\mathbf{d}^{\rm eff}_{\rm I}\cdot(\mathbf{L}_{\rm I}\times\hat{\mathbf{z}})-\mathbf{d}^{\rm eff}_{\rm II}\cdot(\mathbf{L}_{\rm II}\times\hat{\mathbf{z}}))L_z
\end{eqnarray}
while the layer magnetization is given by
\begin{equation}
M_{\alpha}^z\propto \mathbf{d}^{\rm eff}_{\alpha}\cdot (\mathbf{L}_{\alpha} \times\hat{\mathbf{z}})
\end{equation}
Where $\mathbf{d}^{\rm eff}_{\rm I} \equiv \mathbf{d}^{\rm eff}_{\rm 1}$ and $\mathbf{d}^{\rm eff}_{\rm II}\equiv \mathbf{d}^{\rm eff}_{\rm 2}$. We can thus interpret $\mathbf{d}^{\rm eff}_{\alpha}$ as an effective Dzyaloshinskii-Moriya (DM) vector for layer $\alpha$. We will show in the next section that $\mathbf{d}^{\rm eff}_{\alpha}$ originates from the transverse component of the DM interaction and off-diagonal elements of the single-ion anisotropy (SIA) tensor which are both induced by the trimerization distortion.

\subsection{Spin-Lattice coupling from the Dzyaloshinskii-Moriya interaction and Single-Ion Anisotropy}
%

\begin{figure}[t]
\begin{center}
\includegraphics[scale=0.29]{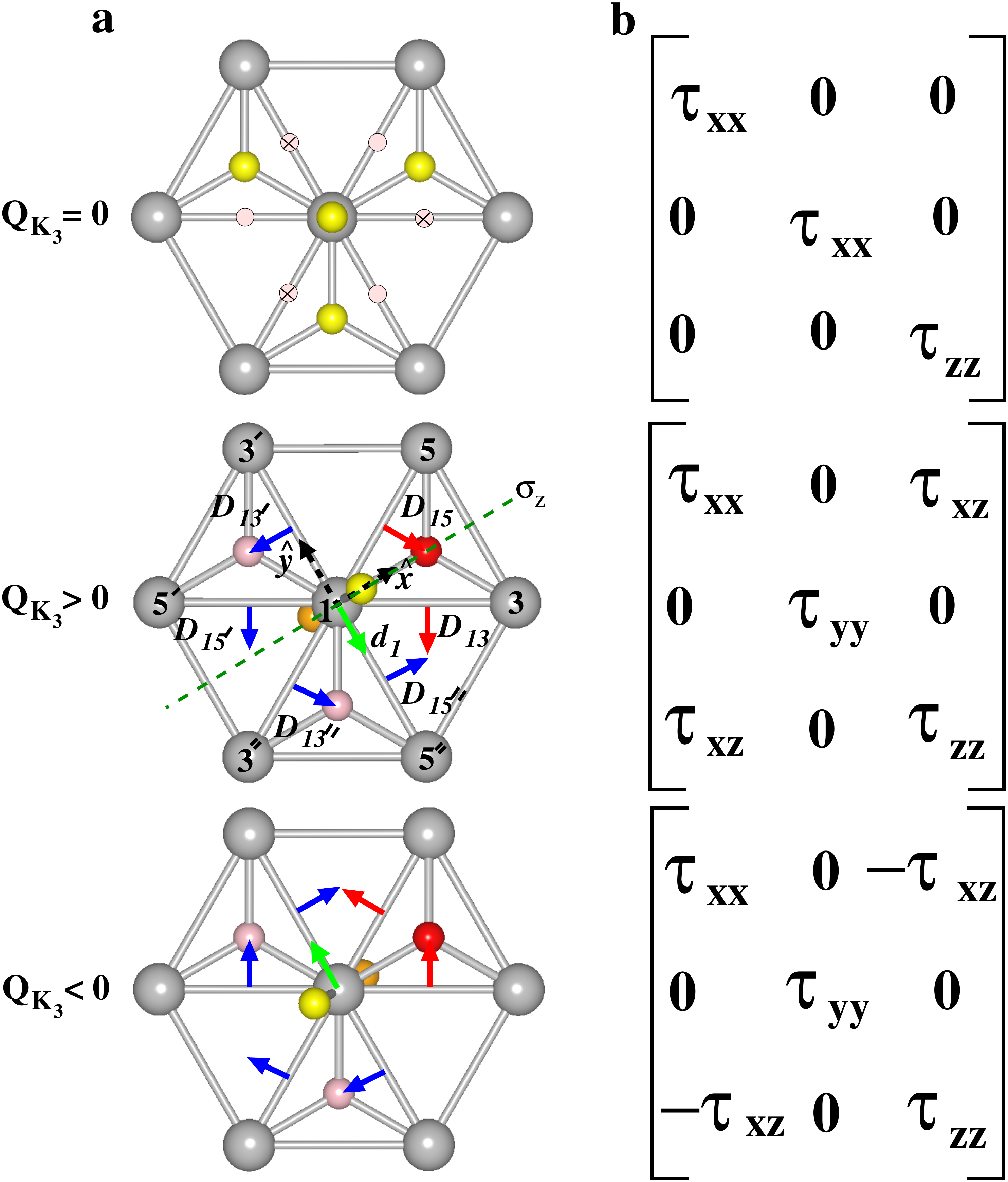}
\end{center}
\caption{\textbf{Dzyaloshinskii-Moriya interactions and Single-Ion Anisotropy.} \textbf{a} The in-plane nn DM vectors for a single triangular layer of TM ions, \textbf{b} SIA tensors for the paraelectric phase and for two opposite directions of the trimer distortion in the ferroelectric phase. All six DM vectors acting on TM site 1 and SIA tensor for this site are shown. DM vectors and SIA tensors for other bonds and sites can be generated by applying the appropriate symmetry operations of crystal space group. In the paraelectric phase only the $z$ component of the nn DM vectors are non-zero since triangular layers are mirror planes. The cross (dot) mark represents direction of a DM vector along positive (negative) $\hat{z}$ axis.
In the ferroelectric structure, the trimer distortion lowers the symmetry, leading to two nonequivalent types of in-plane nn DM vectors: one mediated by TM-O$_p^1$-TM path and the other mediated by TM-O$_p^2$-TM path.}
\label{DMSIA}
\end{figure}

%
\textbf{The Dzyaloshinskii-Moriya interaction of a single layer of bipyramids:}
We considered only in-plane nearest-neighbor (nn) DM interactions which are mediated by TM-O$_{\text{eq}}$-TM paths (here O$_{\text{eq}}$ denotes an equatorial oxygen atom). In the paraelectric phase all nn DM vectors are equivalent and only their $z$ components are nonzero. Note that the $\sigma_{z}$ mirror plane requires DM vectors for adjacent TM-TM bonds to be opposite, see Fig. \ref{DMSIA}a. Physically this is a result of different chiralities of TM-O$_{\text{eq}}$-TM hopping paths for these DM vectors. The $z$ components of the DM vectors confine the spins within the $xy$ plane and don't contribute to canting. They are thus ignored in the following discussion.

In the ferroelectric phase we have two nonequivalent equatorial oxygens: O$^1_{\text{eq}}$ and O$^2_{\text{eq}}$. Consequently, the nn DM vectors split into two nonequivalent types: one mediated by a TM-O$^1_{\text{eq}}$-TM path and the other mediated by a TM-O$^2_{\text{eq}}$-TM path. In addition, both types of DM vectors acquire a nonzero transverse ($xy$) component which is perpendicular to the corresponding TM-TM bond. Note that for DM vectors mediated by TM-O$^2_{\text{eq}}$-TM path the component parallel to the TM-TM bond is, in general, allowed by symmetry due to different orientations of apical oxygen displacements for the two TM ions. However, since apical oxygens have minor effect on TM-TM hopping, the parallel component is small which was confirmed by our first principles calculations. In the following we thus neglected the parallel component to obtain clearer picture of DM interactions. We point out, however, that the inclusion of the parallel component doesn't affect our main conclusions. Transverse components of DM vectors between TM site 1 (see Fig. \ref{DMSIA}a) and its nearest neighbors for different trimer domains are shown in Table ~\ref{table-dm}.

\begin{table*}
\begin{center}
\caption{Transverse components of DM vectors between TM site 1 (see Fig. \ref{DMSIA}a) and its nearest neighbors for different trimer domains.}
\begin{tabular}{|c|c|c|c|}
\hline
$\Phi$ &$\mathbf{D}_{13}^{xy}$ &$\mathbf{D}_{13'}^{xy}$  &$\mathbf{D}_{13''}^{xy}$ \\
\hline
$0$ &$D_{xy}(-\frac{1}{2},-\frac{\sqrt{3}}{2})$ &$D^{\prime}_{xy}(-1,0)$ &$D^{\prime}_{xy}(\frac{1}{2},-\frac{\sqrt{3}}{2})$  \\
\hline
$\pi/3$ &$D^{\prime}_{xy}(-\frac{1}{2},-\frac{\sqrt{3}}{2})$ &$D^{\prime}_{xy}(1,0)$ &$D_{xy}(\frac{1}{2},-\frac{\sqrt{3}}{2})$  \\
\hline
$2\pi/3$ &$D^{\prime}_{xy}(\frac{1}{2},\frac{\sqrt{3}}{2})$ &$D_{xy}(1,0)$ &$D^{\prime}_{xy}(\frac{1}{2},-\frac{\sqrt{3}}{2})$  \\
\hline
$\pi$ &$D_{xy}(\frac{1}{2},\frac{\sqrt{3}}{2})$ &$D^{\prime}_{xy}(1,0)$ &$D^{\prime}_{xy}(-\frac{1}{2},\frac{\sqrt{3}}{2})$  \\
\hline
$4\pi/3$ &$D^{\prime}_{xy}(\frac{1}{2},\frac{\sqrt{3}}{2})$ &$D^{\prime}_{xy}(-1,0)$ &$D_{xy}(-\frac{1}{2},\frac{\sqrt{3}}{2})$  \\
\hline
$5\pi/3$ &$D^{\prime}_{xy}(-\frac{1}{2},-\frac{\sqrt{3}}{2})$ &$D_{xy}(-1,0)$ &$D^{\prime}_{xy}(-\frac{1}{2},\frac{\sqrt{3}}{2})$  \\
\hline
\hline
$\Phi$ &$\mathbf{D}_{15}^{xy}$ &$\mathbf{D}_{15'}^{xy}$ &$\mathbf{D}_{15''}^{xy}$ \\
\hline
$0$  &$D_{xy}(\frac{1}{2},-\frac{\sqrt{3}}{2})$ &$D^{\prime}_{xy}(-\frac{1}{2},-\frac{\sqrt{3}}{2})$ &$D^{\prime}_{xy}(1,0)$ \\
\hline
$\pi/3$ &$D^{\prime}_{xy}(\frac{1}{2},-\frac{\sqrt{3}}{2})$ &$D^{\prime}_{xy}(\frac{1}{2},\frac{\sqrt{3}}{2})$ &$D_{xy}(1,0)$ \\
\hline
$2\pi/3$ &$D^{\prime}_{xy}(-\frac{1}{2},\frac{\sqrt{3}}{2})$ &$D_{xy}(\frac{1}{2},\frac{\sqrt{3}}{2})$ &$D^{\prime}_{xy}(1,0)$ \\
\hline
$\pi$ &$D_{xy}(-\frac{1}{2},\frac{\sqrt{3}}{2})$ &$D^{\prime}_{xy}(\frac{1}{2},\frac{\sqrt{3}}{2})$ &$D^{\prime}_{xy}(-1,0)$ \\
\hline
$4\pi/3$ &$D^{\prime}_{xy}(-\frac{1}{2},\frac{\sqrt{3}}{2})$ &$D^{\prime}_{xy}(-\frac{1}{2},-\frac{\sqrt{3}}{2})$ &$D_{xy}(-1,0)$ \\
\hline
$5\pi/3$ &$D^{\prime}_{xy}(\frac{1}{2},-\frac{\sqrt{3}}{2})$ &$D_{xy}(-\frac{1}{2},-\frac{\sqrt{3}}{2})$ &$D^{\prime}_{xy}(-1,0)$ \\
\hline
\end{tabular}
\end{center}
\label{table-dm}
\end{table*}

Here we derive the relationship between the local structural distortions and the induced $\mathbf{D}^{xy}_{ij}$. Let's consider the layer $\alpha=$ I.  The DM interaction energy (per spin) is given by (see notation in Fig. \ref{DMSIA}a)
\begin{widetext}
\begin{eqnarray}
E_{\rm I}^{\text{DM}}&=&\frac{2}{3}(\mathbf{D}_{13}^{xy}+\mathbf{ D}_{13'}^{xy}+\mathbf{ D}_{13''}^{xy})\cdot(\mathbf{S}_1\times\mathbf{S}_3)+\frac{2}{3}(\mathbf{ D}_{35}^{xy}+\mathbf{ D}_{35''}^{xy}+\mathbf{ D}_{3'5}^{xy})\cdot(\mathbf{S}_3\times\mathbf{S}_5)\nonumber
+\frac{2}{3}(\mathbf{ D}_{51}^{xy}+\mathbf{ D}_{5'1}^{xy}+\mathbf{ D}_{5''1}^{xy})\cdot(\mathbf{S}_5\times\mathbf{S}_1) \\
&= &\frac{2}{3}\mathbf{\bar D}_{13}\cdot(\mathbf{S}_1\times\mathbf{S}_3)+\frac{2}{3}\mathbf{\bar D}_{35}\cdot(\mathbf{S}_3\times\mathbf{S}_5)+\frac{2}{3}\mathbf{\bar D}_{51}\cdot(\mathbf{S}_5\times\mathbf{S}_1)
\end{eqnarray}
where we defined the bar DM vectors
\begin{eqnarray}
\mathbf{\bar D}_{13}&\equiv&\mathbf{ D}_{13}^{xy}+\mathbf{ D}_{13'}^{xy}+\mathbf{ D}_{13''}^{xy}=\bar D_{xy}\left(\cos(\Phi-2\pi/3),\sin(\Phi-2\pi/3)\right) \\
\mathbf{\bar D}_{35}&\equiv&\mathbf{ D}_{35}^{xy}+\mathbf{ D}_{35''}^{xy}+\mathbf{ D}_{3'5}^{xy}=\bar D_{xy}\left(\cos(\Phi),\sin(\Phi)\right) \\
\mathbf{\bar D}_{51}&\equiv&\mathbf{ D}_{51}^{xy}+\mathbf{ D}_{5'1}^{xy}+\mathbf{ D}_{5''1}^{xy})=\bar D_{xy}\left(\cos(\Phi+2\pi/3),\sin(\Phi+2\pi/3)\right)
\end{eqnarray}
\end{widetext}
with $\bar D_{xy}=D_{xy}+D^{\prime}_{xy}$. The above expressions can be easily obtained from Table \ref{table-dm} or Fig. \ref{DMSIA}a. Note that the bar DM vectors have magnitude $\bar D_{xy}$ and form a 120$^\circ$ angle with each other.

\begin{figure}
\begin{center}
\includegraphics[scale=0.3]{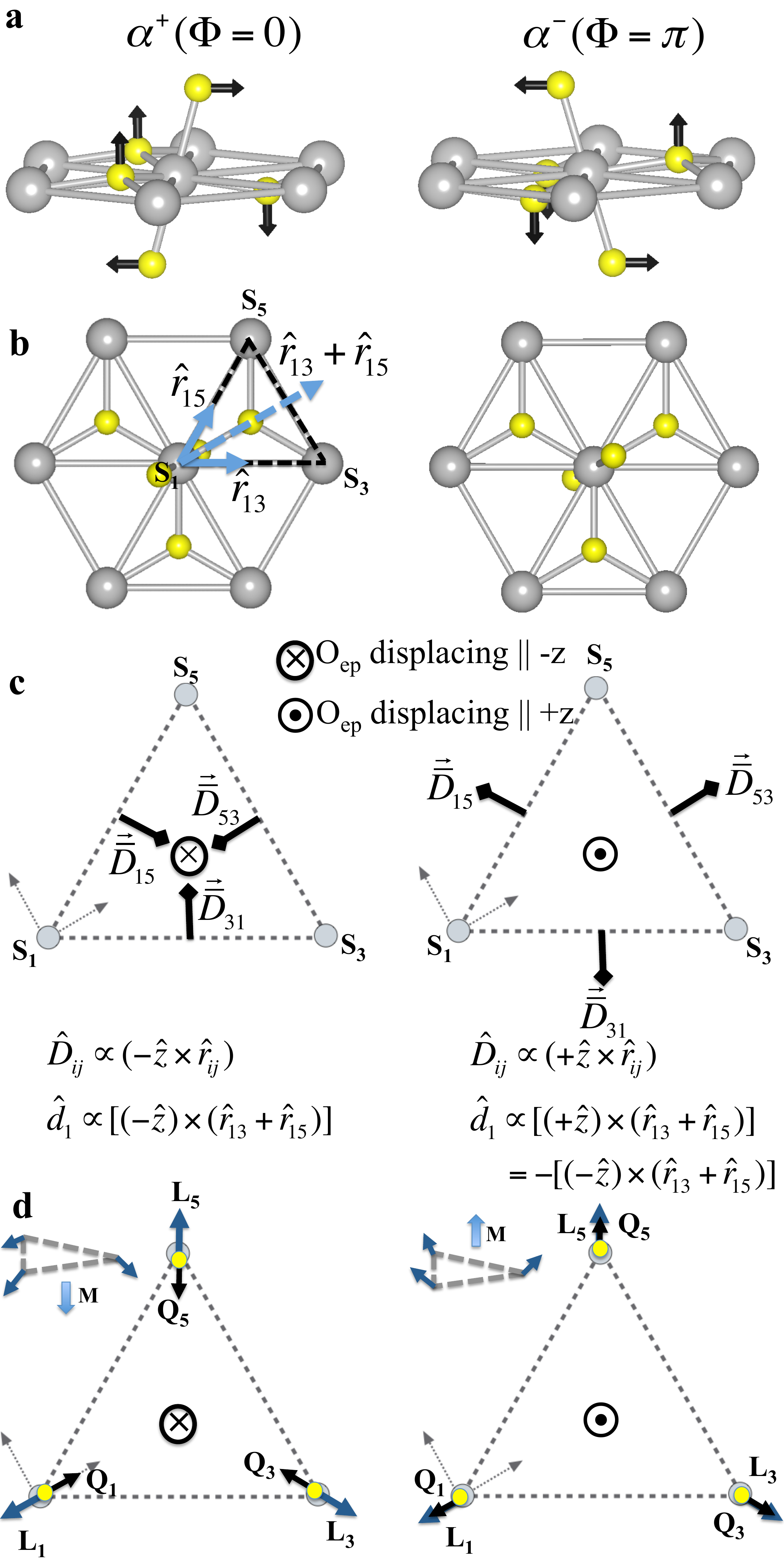}
\end{center}
\caption{\textbf{Microscopic: Basic ingredients to describe connection between microscopic spin-lattice model and a simple phenomenological model.}
\textbf{a} and \textbf{b}, local trimer distortion around $\textbf{S}_1$ from two different projections.
\textbf{c}, the transverse components of DM vectors are shown for a single triangle in a single layer. Cross and dot signs denote displacement of the equatorial oxygen along -$\hat{z}$ and +$\hat{z}$ respectively.
\textbf{d}, demonstrates the relation $\textbf{M}_z \propto (\textbf{L}.\textbf{Q}) \hat{z}$.}
\label{Fig3_supp}
\end{figure}

Therefore, a much simpler picture emerges;  the relationship between the local structural distortions and the induced $\mathbf{D}^{xy}_{ij}$ can be derived by considering a single triangle of spins ($\mathbf{S}_1$, $\mathbf{S}_3$ and $\mathbf{S}_5$) interacting by the bar DM vectors, see Fig.~\ref{Fig3_supp} for the case of the  $\alpha^+$ and $\alpha^-$ domains.

Since all spins cant in the same direction we can write $\mathbf{S}_i = \mathbf{L}_i + \mathbf{M}_{\rm I}$ where $\mathbf{L}_i$ are defined in Fig. 1d of the main manuscript and $\mathbf{M}_{\rm I}$ is the parallel to the $z$ axis layer magnetization. We then obtain

\begin{widetext}
\begin{equation}
E_{\rm I}^{\text{DM}}= \frac{2}{3}(\mathbf{\bar D}_{13} +  \mathbf{\bar D}_{15})   \cdot  (\mathbf{L}_1 \times \mathbf{M}_{\rm I})
+ \frac{2}{3}(\mathbf{\bar D}_{35} +  \mathbf{\bar D}_{31})   \cdot  (\mathbf{L}_3 \times \mathbf{M}_{\rm I})
+ \frac{2}{3}(\mathbf{\bar D}_{51} +  \mathbf{\bar D}_{53})   \cdot  (\mathbf{L}_5 \times \mathbf{M}_{\rm I})=\frac{1}{3}\sum_{i=1,3,5} \mathbf{d}_i\cdot [ \mathbf{L}_i\times  \mathbf{M}_{\rm I}]
\end{equation}

where the $\mathbf{d}_{i}$'s, e.g., $\mathbf{d}_{1} \equiv  2(\mathbf{\bar D}_{13}+ \mathbf{\bar D}_{15})$, are the effective, transverse DM-interactions. Using $A\cdot (B\times C) = B\cdot (C\times A)$ and $\mathbf{M}_{\rm I}= M^z_{\rm I} \mathbf{\hat{z}}$ the DM energy can be rewritten as

\begin{equation}
E_{\rm I}^{\text{DM}} =\frac{1}{3}M_{\rm I}^z\sum_{i=1,3,5} \mathbf{L}_i\cdot [ \mathbf{\hat{z}}\times \mathbf{d}_{i}] = \frac{2\sqrt{3}}{3}\bar{D}_{xy}M_{\rm I}^z\sum_{i=1,3,5}\mathbf{L}_{i}\cdot\mathbf{\hat Q}_i
\label{E_DM}
\end{equation}
\end{widetext}
where we used $2\sqrt{3}\bar{D}_{xy}\mathbf{\hat{Q}}_i=\mathbf{\hat{z}}\times \mathbf{d}_i$ with $\mathbf{\hat{Q}}_i$'s being unit vectors defined in Fig. 1d of the main manuscript. The relation between $\mathbf{d}_i$ and $\mathbf{\hat{Q_i}}$ can be straightforwardly obtained from Eqs. \ref{E_DM} but more physical insight into this relation can be gained by noting that $\mathbf{D}_{ij}^{xy}\propto \mathbf{\hat{r}}_{ij}\times\mathbf{u}_{\text{O}_{\text{eq}}}$ where $\mathbf{\hat{r}}_{ij}$ is the unit vector pointing from site $i$ towards $j$ and $\mathbf{u}_{\text{O}_{\text{eq}}}$ is the displacement of the equatorial oxygen away from the plane (e.g., due to the tilting of the bipyramid) which is zero in the PE phase and parallel to the $z$ axis in the FE phase. For the $\alpha^{\pm}$ trimer domains we find
\begin{equation}
\mathbf{d}_{1} \propto \pm\mathbf{\hat{z}}\times(\mathbf{\hat{r}}_{13}+\mathbf{\hat{r}}_{15})
\end{equation}
As seen from Fig. \ref{Fig3_supp}b, $\mathbf{\hat{r}}_{13}+\mathbf{\hat{r}}_{15}= \mathbf{\hat{Q}}_1$ giving

 \begin{equation}
\mathbf{\hat{Q}}_1 \propto \pm\mathbf{\hat{z}}\times\mathbf{d}_{1}\end{equation}

Other   $\mathbf{d}_{i}$ can be found by cyclic permutations: $1\rightarrow3$, $3\rightarrow5$, and $5\rightarrow1$. The same results can be obtained for other trimer domains.

Note that

\begin{equation}
\mathbf{L}_{1}\cdot\mathbf{\hat Q}_1=\mathbf{L}_{3}\cdot\mathbf{\hat Q}_3=
\mathbf{L}_{5}\cdot\mathbf{\hat Q}_5 \equiv \mathbf{L}_{\rm I}\cdot\mathbf{\hat Q}_{\rm I} =L_{xy}\cos(\Phi-\Psi_1)
\end{equation}

We thus obtain

\begin{equation}
E_{\rm I}^{\text{DM}} = 2\sqrt{3}\bar{D}_{xy}M_{\rm I}^z \mathbf{L}_{\rm I}\cdot\mathbf{\hat Q}_{\rm I} =  M_{\rm I}^z\mathbf{L}_{\rm 1}\cdot\mathbf{\hat{z}}\times \mathbf{d}_{\rm 1}
\end{equation}

which leads to

\begin{equation}
M_{\rm I}^z \propto \mathbf{L}_1\cdot\mathbf{\hat{z}}\times \mathbf{d}_{1} = 2\sqrt{3}\bar{D}_{xy}\mathbf{L}_{\rm I}\cdot\mathbf{\hat Q}_{\rm I}
\end{equation}

in agreement with Eqs.\ref{Mz_Landau} obtained from the Landau theory. It is now clear that the microscopic origin of the layer magnetization is the the trimer induced transverse components of the DM interactions which cant spins away from the $xy$ plane. As we will see in the next section, however, there is also another contribution to the canting that originates from the single-ion anisotropy (SIA).

%

%
\textbf{The Single-ion anisotropy of a single layer of bipyramids:}
In the paraelectric phase the crystal field has the same orientation for all TM ions so the single-ion anisotropy (SIA) tensor, $\mathbf{\hat\tau}_i$, does not depend on magnetic site index $i$. A global coordinate system (see Fig. \ref{DMSIA}) can be thus chosen in which $\mathbf{\hat\tau}$ is diagonal with elements $\tau^{xx}$, $\tau^{yy}$, $\tau^{zz}$. A uniaxial site symmetry and the zero-trace condition lead to  $\tau^{xx}=\tau^{yy}=-\tau^{zz}/2$.

On the other hand, in the ferroelectric phase the crystal field may have different orientations for different TM ions and therefore $\mathbf{\hat\tau}_i$ does depend on $i$. Even though all TM ions remain equivalent and thus SIA tensors for different magnetic sites are related by symmetry, in any global coordinate system SIA tensor for some magnetic ions have off diagonal components. In addition, the uniaxial site symmetry is lost in the ferroelectric phase leading to the in-plane anisotropy ($\tau^{xx}\neq\tau^{yy}$). In the coordinate system as in Fig. \ref{DMSIA}a the SIA tensor for site 1 in the $\Phi=0$ trimer domain is given by

\begin{equation}
\mathbf{\hat\tau} =\begin{bmatrix}
\tau_{xx}&0&\tau_{xz}\\
0&\tau_{yy}&0\\
\tau_{xz}&0&\tau_{zz}\\
\end{bmatrix}
\label{SIA}
\end{equation}

For a general trimer domain the SIA tensor for site 1 is given by $R_{\Phi}\mathbf{\hat\tau}R^{-1}_{\Phi}$ where $R_{\Phi}$ is a rotation matrix

\begin{equation}
R_{\Phi} =\begin{bmatrix}
\cos\Phi&-\sin\Phi&0\\
\sin\Phi&\cos\Phi&0\\
0&0&1\\
\end{bmatrix}
\end{equation}

The effect of trimer distortion on the components of the SIA tensor can be understood if we assume that the crystal field for a given TM ion is determined solely by its oxygen bypyramid. In this case the components of $\mathbf{\hat\tau}$ in Eq. (\ref{SIA}) can be expressed in terms of the tilting angle $\theta$ and the value of $\tau_{zz}$ in the paraelectric phase (hereafter denoted by $\tau^0_{zz}$). For site 1 we have:

\begin{align}
\tau_{xx}=-\tau^0_{zz}/2 \\
\tau_{yy}=-\cos^2(\theta)\tau^0_{zz}/2 + \sin^2(\theta)\tau^0_{zz} \\
\tau_{zz}=\cos^2(\theta)\tau^0_{zz} - \sin^2(\theta)\tau^0_{zz}/2 \\
\tau_{xz}=-3\sin(2\theta)\tau^0_{zz}/4
\label{SIA0}
\end{align}

First principles calculations show that the in-plane anisotropy is very small (see Table.~\ref{parameters2}). Indeed, as seen from Eqs. (\ref{SIA0}) this difference is proportional to $\sin^2\theta$ which is a very small quantity.  On the other hand, the off-diagonal component, $\tau_{xz}$ is proportional to $\sin2\theta$ and is correspondingly substantially larger and plays important role in the canting.

Let's consider the SIA contribution to the canting energy (per spin) for the layer $\alpha=$ I
\begin{widetext}
\begin{equation}
E_{\rm I}^{SIA}=\frac{1}{3}\sum_{i=1,3,5}\mathbf{S}_i\cdot\mathbf{\hat \tau}_1\cdot\mathbf{S}_i =\frac{2}{3}\tau_{xz}M_{\rm I}^z\sum_{i=1,3,5}\mathbf{L}_i\cdot\mathbf{\hat Q}_i=2\tau_{xz}M_{\rm I}^z\mathbf{L}_{\rm I}\cdot\mathbf{\hat Q}_{\rm I}
\end{equation}
\end{widetext}
where we kept only the terms proportional to $M_{\rm I}^z$. The above equation has a similar form as Eq. \ref{E_DM}. Indeed, we can define a DM-like vector $\mathbf{d}_{i}^{SIA}$ as $2\tau_{xz}\mathbf{\hat Q}_{i}=\mathbf{\hat z}\times\mathbf{d}_{i}^{SIA}$. We then get
\begin{equation}
E_{\rm I}^{\rm SIA} = \frac{1}{3}\sum_{i=1,3,5} \mathbf{d}_{i}^{SIA}\cdot [ \mathbf{L}_i\times  \mathbf{M}_{\rm I}^z]
\label{dSIA}
\end{equation}

\textbf{The total energy and magnetization of a single layer of bipyramids:}
Combining Eqs. (\ref{E_DM}) and (\ref{dSIA}) we obtain
\begin{equation}
E_{\rm I}^{\rm canting}=E_{\rm I}^{\rm DM}+E_{\rm I}^{\rm SIA} = \frac{1}{3}\sum_{i=1,3,5} \mathbf{d}^{\rm eff}_{i}\cdot [ \mathbf{L}_i\times  \mathbf{M}_{\rm I}]
\label{dtotal}
\end{equation}
where $\mathbf{d}^{\rm eff}_{i}=\mathbf{d}_{i}+\mathbf{d}_{i}^{SIA}$ is the effective transverse DM vector with the magnitude $|\mathbf{d}^{\rm eff}|  = 2|\sqrt{3}\bar D_{xy}+\tau_{xy}|$. 

Again using $A\cdot (B\times C) = B\cdot (C\times A)$ and $\textbf{M}_{\rm I}= M_{\rm I}^z \mathbf{\hat{z}}$, $E_{\rm canting}$ can be rewritten as
\begin{eqnarray}
E_{\rm I}^{\rm canting} &=&M_{\rm I}^z\sum_{i=1,3,5} \mathbf{L}_i\cdot [ \mathbf{\hat z}\times  \mathbf{d}^{\rm eff}_{i}] \nonumber\\
&=& |\mathbf{d}^{\rm eff}|M_{\rm I}^z\mathbf{L}_{\rm I}\cdot  \mathbf{\hat Q}_{\rm I}\end{eqnarray}
so that the layer magnetization due to canting is given by
\begin{equation}
M_{\rm I}^z\approx\frac{(\sqrt{3}\bar D_{xy}+\tau_{xz})}{6J_{nn}}\,\,\mathbf{L}_{\rm I}\cdot\mathbf{Q}_{\rm I}
\end{equation}

where $J_{nn}$ is the nn exchange interaction.

We thus recovered the result from Landau Theory.

\textbf{The real  structure: the stacking of two layers:}
Let us consider now a real hexa structure which is composed of two layers $\alpha=$ I and $\alpha=$ II, each with a, in principle different, layer magnetization, $M^z_{\rm I}$ and $M^z_{{\rm II}}$ respectively. The canting energy is

\begin{widetext}
\begin{equation}
E_{\rm canting}=(E_{\rm I}^{\rm canting}+E_{\rm II}^{\rm canting})/2=  |\mathbf{d}^{\rm eff}| \left[ M^z_{\rm I} (\textbf{L}_{\rm I}\cdot\textbf{Q}_{\rm I}) + M^z_{\rm II}(\textbf{L}_{\rm II}\cdot\textbf{Q}_{\rm II})\right]/2
\label{DMfinal}
\end{equation}
\end{widetext}
This result shows clearly that the B$_1$ state displays weak-antiferromagentism, wAFM, i.e., there is a  canting of the spins out of each spin plane, but since the projection  changes sign in adjacent layers, i.e.,  $\textbf{L}_{\rm I}\cdot\textbf{Q}_{\rm I}$ = -$\textbf{L}_{\rm II}\cdot\textbf{Q}_{\rm II}$,  no net magnetization exists, ${M}_z = M^z_{\rm I} + M^z_{\rm II}=0$. In the A$_2$ phase, however,  the projection has the same sign in adjacent layers, $\textbf{L}_{\rm I}\cdot\textbf{Q}_{\rm I}$ = +$\textbf{L}_{\rm II}\cdot\textbf{Q}_{\rm II}$,  and therefore $M^z_{\rm I} = M^z_{\rm II}$ leading to a net magnetization along the $z$ axis.
This result explains our first-principles calculations displayed in Fig.1 of the main manuscript and provides a microscopic justification for the results of our simple Landau theory.

The above results can be rewritten in terms of the trimer phase and the spin angles
\begin{eqnarray}
E^{A_2}_{\rm canting}&\propto &{M}_{z}{Q}_{K_3}L_{xy} \left[ {\rm cos}(\Psi_1-\Phi) - {\rm cos}(\Psi_2+\Phi) \right] \label{trilinear2M} \\
 E^{B_1}_{\rm canting}&\propto &{L}_{z} {Q}_{K_3}L_{xy} \left[ {\rm cos}(\Psi_1-\Phi) + {\rm cos}(\Psi_2+\Phi) \right]
\label{trilinear2L}
\end{eqnarray}
describing weak-ferromagnetism for the $A_2$ phase and weak-antiferromagnetism for the $B_1$ phase respectively.
Notice that  if $\textbf{P}$ switched via rotating $\Phi$ by $|\pi/3|$, e.g.,    $\alpha^+\rightarrow\beta^-$,  $\textbf{L}$ must rotate by either $|\pi/3|$ ($\textbf{L}\cdot\textbf{Q}_{\alpha^+} = -1 \rightarrow\textbf{L}\cdot\textbf{Q}_{\beta^-} =-1$) or $|2\pi/3|$,  ($\textbf{L}\cdot\textbf{Q}_{\alpha^+} = -1 \rightarrow\textbf{L}\cdot\textbf{Q}_{\beta^-} = +1$). \\

%
\subsection{Magnetoelectric effect}
%

The ferroelectric phase in the A$_2$ magnetic structure has P6$_3$c'm' space group. The corresponding point group is 6m'm' which allows for magnetoelectric (ME) effect with magnetoelectric susceptibility tensor,
\begin{equation}
\mathbf{\hat\alpha} =\begin{bmatrix}
\alpha_{\perp}&0&0\\
0&\alpha_{\perp}&0\\
0&0&\alpha_{\parallel}\\
\end{bmatrix}
\end{equation}

In order to understand the origin of this ME coupling we consider Landau expansion with respect to the P6$_3$/mmc1$^{\prime}$ reference structure. The part of the free energy that depends on $M_z$ can be written as
\begin{equation}
F(M_z)=\frac{1}{2}a_MM_z^2-\frac{1}{2}c_{tr}Q_{K_3}M_z(\mathbf{\hat Q}_{\rm I}\cdot \mathbf{L}_{\rm I} + \mathbf{\hat Q}_{\rm II}\cdot\mathbf{L}_{\rm II})
\end{equation}
where we defined $\mathbf{\hat Q}_{\alpha}=\mathbf{Q}_{\alpha}/Q_{K_3}$. Minimizing with respect to $M_z$ we find an equilibrium magnetization

\begin{equation}
M_z=\frac{1}{2}\frac{c_{tr}}{a_M}Q_{K_3}(\mathbf{\hat Q}_{\rm I}\cdot \mathbf{L}_{\rm I} + \mathbf{\hat Q}_{\rm II}\cdot\mathbf{L}_{\rm II})
\label{M_z}
\end{equation}
Assuming the in-plane spin components are rigid (this assumptions is rigorous in the A$_2$ phase) the $zz$ component of the ME susceptibility is
\begin{widetext}
\begin{equation}
\alpha_{\parallel}=\left.\frac{\partial M_z}{\partial E_z}\right|_{E_z=0}=\frac{1}{2}\frac{c_{tr}}{a_M}(\mathbf{\hat Q}_{\rm I}\cdot \mathbf{L}_{\rm I} + \mathbf{\hat Q}_{\rm II}\cdot\mathbf{L}_{\rm II})\left.\frac{\partial Q_{K_3}}{\partial E_z}\right|_{E_z=0}
\label{me2}
\end{equation}
In order to find $\left.\frac{\partial Q_{K_3}}{\partial E_z}\right|_{E_z=0}$ we consider the free energy as a function of $P_z$ and $Q_{K_3}$
\begin{equation}
F(P_z,Q_{K_3})=\frac{1}{2}a_PP_z^2+\frac{1}{2}a_QQ_{K_3}^2+\frac{1}{4}b_QQ_{K_3}^4-dP_zQ_{K_3}^3\cos{3\Phi}+\frac{1}{2}d'P_z^2Q_{K_3}^2-E_zP_z
\label{fen}
\end{equation}
In above $M_z$ was integrated out resulting in renormalization of the $a_Q$ coefficient. Minimizing with respect to $P_z$ we obtain
\begin{equation}
P_z=\frac{dQ_{K_3}^3\cos{3\Phi}+E_z}{a_P+d'Q_{K_3}^2}
\end{equation}
We assume that we are well below the trimerization transition and $Q_{K_3}$ is large and satisfies $\frac{d'}{a_P}Q_{K_3}^2>>1$. Then the above equation simplifies to
\begin{equation}
P_z\approx\frac{d}{d'}Q_{K_3}\cos{3\Phi} +\frac{1}{d'Q_{K_3}^2}E_z
\label{Plin}
\end{equation}
Minimization of Eq. (\ref{fen}) with respect to $Q_{K_3}$ leads to
\begin{equation}
a_QQ_{K_3}+b_QQ_{K_3}^3-3dP_zQ_{K_3}^2\cos{3\Phi}+d'P_z^2Q_{K_3}=0
\label{Qeq}
\end{equation}
Substituting (\ref{Plin}) into (\ref{Qeq}) we obtain
\begin{equation}
a_QQ_{K_3}+\tilde b_QQ_{K_3}^3-\frac{d}{d'}E_z\cos{3\Phi}+O(E_z^2)=0
\label{Qeq2}
\end{equation}
where $\tilde b_Q=b_Q-2d^2/d'$ and we took into account that within any trimer domain $\cos{3\Phi}=\pm1$. Taking derivative with respect to $E_z$ at $E_z=0$ we obtain
\begin{equation}
a_Q\left.\frac{\partial Q_{K_3}}{\partial E_z}\right|_{E_z=0}+3\tilde b_QQ_{K_3}^2(E_z=0)\left.\frac{\partial Q_{K_3}}{\partial E_z}\right|_{E_z=0}-\frac{d}{d'}\cos{3\Phi}=0
\end{equation}
\end{widetext}
From (\ref{Qeq2}) we obtain $Q_{K_3}^2(E_z=0)=-a_Q/\tilde{b}_Q$ leading to
\begin{equation}
\left.\frac{\partial Q_{K_3}}{\partial E_z}\right|_{E_z=0}=-\frac{1}{2a_Q}\frac{d}{d'}\cos{3\Phi}
\label{dQdE}
\end{equation}
Therefore, the magnetoelectric susceptibility becomes
\begin{equation}
\alpha_{\parallel}=-\frac{1}{4}\frac{c_{tr}d}{a_Ma_Qd'}(\mathbf{\hat Q}_{\rm I}\cdot \mathbf{L}_{\rm I} + \mathbf{\hat Q}_{\rm II}\cdot\mathbf{L}_{\rm II})\cos{3\Phi}
\label{me2}
\end{equation}
Note that from (~\ref{M_z}) and (~\ref{Plin}) it follows that,
\begin{equation}
\alpha \propto \dfrac{\partial M_z}{\partial Q_{K_3}}*\dfrac{\partial P_z}{\partial Q_{K_3}}\equiv \partial_Q M_z* \partial_Q P_z
\end{equation}

Few comments are in order. First, $\alpha_{\parallel}$ is nonzero only when $(\mathbf{\hat Q}_{\rm I}\cdot \mathbf{L}_{\rm I} + \mathbf{\hat Q}_{\rm II}\cdot\mathbf{L}_{\rm II})$ is nonzero which is exactly the condition for existence of weak ferromagnetism that requires that the magnetic configuration has a nonzero A$_2$ component. Second, if $(\mathbf{\hat Q}_{\rm I}\cdot \mathbf{L}_{\rm I} + \mathbf{\hat Q}_{\rm II}\cdot\mathbf{L}_{\rm II})$ is fixed (i.e., the projections $\mathbf{\hat Q}_{\alpha}\cdot\mathbf{L}_{\alpha})$ are equal in all domains), then the sign of $\alpha_{\parallel}$ switches as we go from prime ($\Phi=\pi, \pm\pi/3$) to nonprime ($\Phi=0, \pm2\pi/3, $) trimer domains. In other words, the domains with parallel $M_z$ and $P_z$ have an opposite sign of $\alpha_{\parallel}$ than domains with $M_z$ and $P_z$ antiparallel.

\begin{figure}
\begin{center}
\includegraphics[scale=0.26]{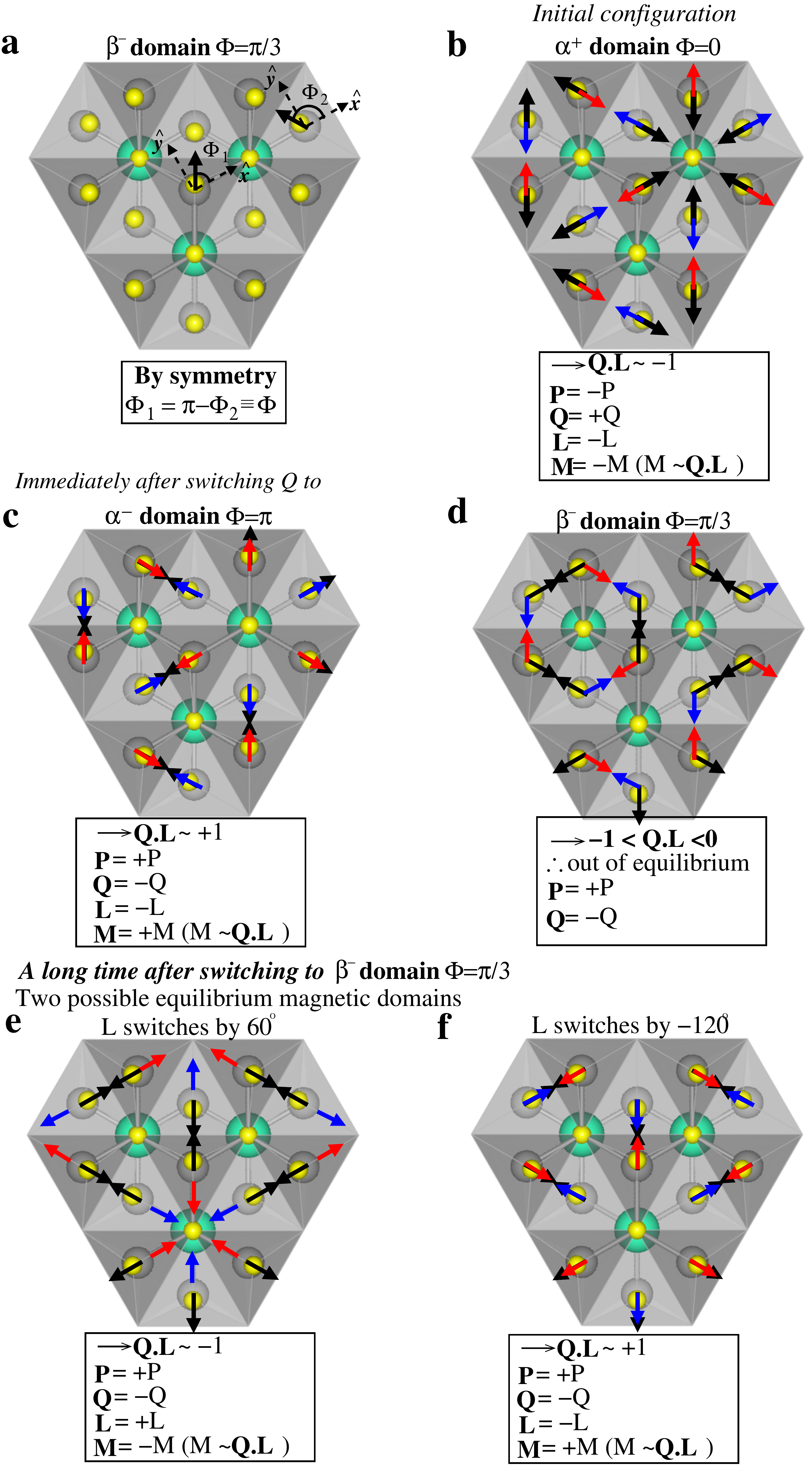}
\end{center}
\caption{\textbf{Bulk cross-coupling of polarization and magnetization in A$_2$ phase.} Thought experiment to elucidate the  cross-coupling of ferroelectricity and magnetism mediated by the trimer distortion for the predicted A$_2$ magnetic ground state. \textbf{a,} Definition of local trimer angles $\Phi_1$ and $\Phi_2$, \textbf{b,} Initial equilibrium  $\alpha^+$ domain $\equiv$ \{-P,+Q, -L$_{xy}$, -M$_z$\}. Immediately after switching \textbf{c,} to the  $\alpha^-$  domain, $L_{xy}$ still in ground state, therefore -M$_z$ $\rightarrow$ +M$_z$;  $\alpha^-$ domain $\equiv$ \{+P,-Q, -L$_{xy}$, +M$_z$\}, \textbf{d,}   $\beta^-$  domain, $L_{xy}$ is not in a ground state and must rotate either, \textbf{e,}  60$^{\circ}$, therefore -M$_z$ $\rightarrow$ -M$_z$; $\beta^-$ domain $\equiv$ \{+P,-Q, +L$_{xy}$, -M$_z$\}, or   \textbf{f,}  120$^{\circ}$,  therefore -M$_z$ $\rightarrow$ +M$_z$; $\beta^-$ domain $\equiv$ \{+P,-Q, -L$_{xy}$, +M$_z$\}. }
\label{Fig4_supp}
\end{figure}

%
\subsection{Implications of the trilinear coupling}
%

The trilinear coupling of Eq.~\ref{DMfinal} is quite remarkable. It implies that in the A$_2$ phase  the trimer distortion not only induces a polarization but also mediates a non-trivial bulk  $\textbf{P}$-$\textbf{M}$  coupling. To make this clear, we  consider a thought experiment in which an electric field, $\mathbf{E}$ applied along the $z$ axis can switch  $\textbf{P}$ to any one of the three trimer domains with $-\textbf{P}$ (more in the Discussion). Let the system be initially in the $\alpha^+$ domain with polarization $+\textbf{P}$  and  ${\mathbf{Q}}_{\alpha}\cdot {\mathbf{L}}_{\alpha} = -1$, Fig.~\ref{Fig4_supp}b. Then there are two possible scenarios:

\{$\alpha^+\rightarrow \alpha^-$\}. In a proper FE like PbTiO$_3$ the structure of the $+\textbf{P}$ domain is related to the  $-\textbf{P}$ domain by  a reversal in  the  direction of the  polar distortions {\it w.r.t} the PE structure, e.g, the Ti$^{4+}$ ion moving from up to down.  The analogous situation in the hexa systems corresponds to a structural change from a  2-up/1-down buckling and  tilting `out' of the R-planes and bypyramids, respectively, to a 1-up/2-down and tilting `in', while remaining in the same distinct domain, e.g., $\alpha^+$. This corresponds to switching $\textbf{P}$  via rotating $\Phi$ by $\pi$ (Fig.~\ref{Fig4_supp}c). In this $\alpha^-$ domain, because of  Eq.~\ref{DMfinal}, either $\mathbf{L}_{\alpha}$  has to rotate 180$^{\circ}$ (${\mathbf{Q}}_{\alpha}\cdot {\mathbf{L}}_{\alpha} = -1$) or the small canting angle has to change sign (${\mathbf{Q}}_{\alpha}\cdot {\mathbf{L}}_{\alpha} = +1$). It is not unreasonable to expect the latter to be more favorable, leading to a reversal of  $\textbf{M}$.

\{$\alpha^+\rightarrow \beta^-$\}. The improper nature of ferroelectricity, however, offers an even more interesting possibility is that there exists three distinct and accessible domains ($\alpha$, $\beta$, and $\gamma$).
As an example let $\textbf{P}$ switch via rotating $\Phi$ by $\pi/3$  and consider the configuration immediately after,  Fig.~\ref{Fig4_supp}d.  In this $\beta^-$ domain $-1<{\mathbf{Q}}_{\alpha}\cdot {\mathbf{L}}_{\alpha} <0$, implying the system is not in equilibrium, and therefore $\mathbf{L}_{\alpha} $ must rotate by either $|\pi/3|$, Fig.~\ref{Fig4_supp}e, or $|2\pi/3|$, Fig.~\ref{Fig4_supp}f. In the former case ${\mathbf{Q}}_{\alpha}\cdot {\mathbf{L}}_{\alpha} =-1$ as in the initial $\alpha^+$ configuration, therefore  $\textbf{M}$  is not reversed, while in the latter ${\mathbf{Q}}_{\alpha}\cdot {\mathbf{L}}_{\alpha} = +1$ and $\textbf{M}$ switches 180$^{\circ}$.

\subsection{Discussion: Possible realizations of predictions}

\begin{figure*}
\begin{center}
\includegraphics[scale=0.6]{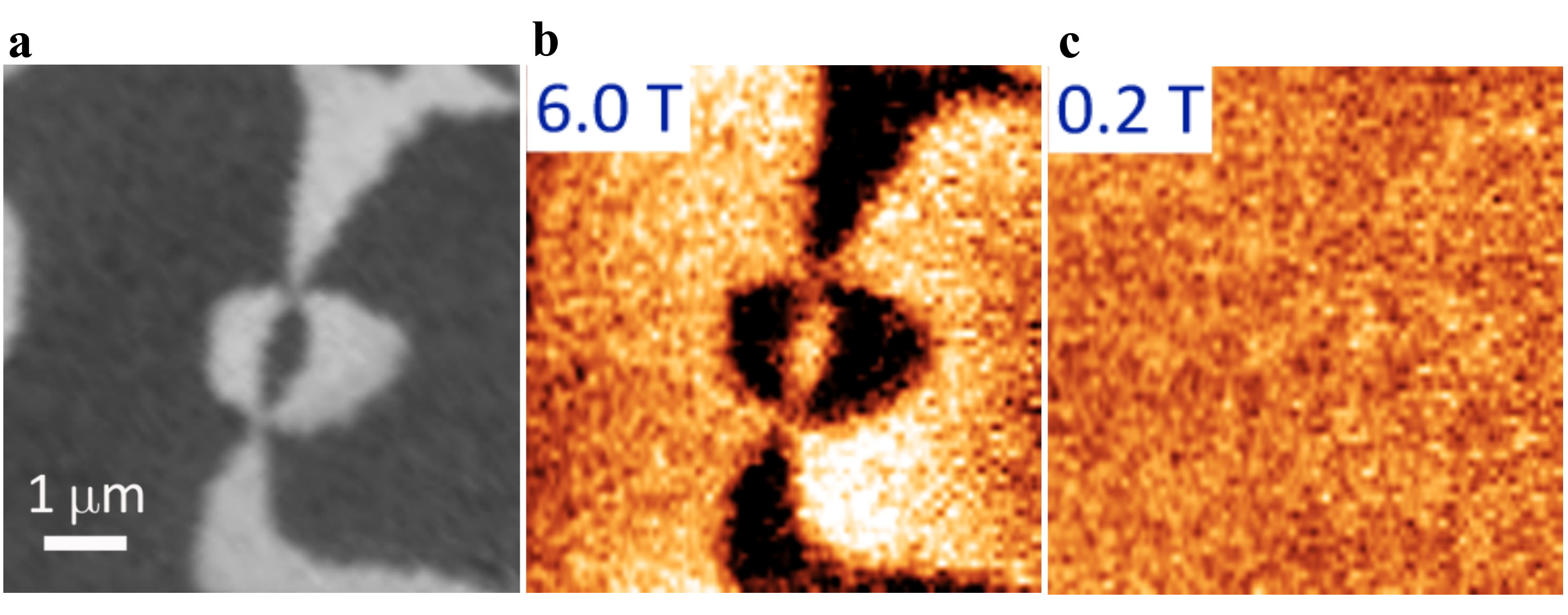}
\end{center}
\caption{\textbf{Magnetoelectric force microscopy (MeFM) images}. \textbf{a} Piezoelectric force microscopy (PFM) image taken under ambient condition. \textbf{b} and \textbf{c}, show MeFM images taken under the application of 6 and 0.2 T external magnetic field, respectively.}
\label{MEEPT}
\end{figure*}

\textbf{Realization 1: The hexagonal Manganites:}
Fig.~\ref{MEEPT} shows Piezoelectric force microscopy (PFM) and Magnetoelectric force microscopy (MeFM) images. The MeFM images, were aligned to PFM image by the topographic landmarks and are in the same color scale, 9.18 mHz.

\textbf{Realization 2: The A$_2$ ground state in hexa RFeO$_3$:}
%
\subsubsection{Origin of ferroelectricity in hexaferrites}
%

Experimental studies of hexagonal ferrites RFeO$_3$ indicates that at room temperature these materials have a polar P6$_3$cm structure \cite{bossak04,magome10} while at high temperature they crystallize in the paraelectric P6$_3$/mmc phase \cite{jeong12a}.According to the group theory analysis there are three possible phase transition sequences connecting the high-temperature phase with the low-temperature P6$_3$cm phase \cite{fennie05}: (1) a direct transition between the two phases by freezing-in a K$_3$ zone boundary phonon mode, (2) a transition to the intermediate P6$_3$mc ferroelectric phase by softening of the $\Gamma^-_2$  zone-center polar mode followed by the transition to the low-temperature phase, and (3) a transition to the intermediate non-polar P6$_3$/mcm phase by freezing-in a zone boundary K$_1$ mode followed by transition to the low temperature phase.

\begin{table}
\caption{Calculated frequencies of relevant phonon modes for the high-temperature phase of RTMO$_3$ (TM=Fe, Mn). In addition, electric polarization ($P$) for the fully relaxed low-temperature structures are listed.}
\begin{center}
\begin{tabular}{c|ccc|c}
\hline
\hline
System&\multicolumn{3}{c|}{$\omega$ (cm$^{-1}$)}& $P_s$ \\
&$\Gamma _2^{-1}$&K$_1$&K$_3$&($\mu$C/cm$^2$)\\
\hline
HoFeO$_3$&73&276&114i&9.0\\
ErFeO$_3$&75&271&113i&9.2\\
TmFeO$_3$&78&268&112i&9.5\\
YbFeO$_3$&102&298&97i&9.6\\
LuFeO$_3$&84&263&108i&9.8\\
\hline
\hline
HoMnO$_3$&65&249&125i&6.9\\
ErMnO$_3$&61&245&128i&7.0\\
TmMnO$_3$&56&240&130i&7.2\\
YbFeO$_3$&92&266&127i&7.4\\
LuFeO$_3$&54&232&132i&7.4\\
\hline
\end{tabular}
\end{center}
\label{phonon}
\end{table}

Calculated phonon frequencies in the high-temperature phase for the above mentioned phonon modes are presented in Table.~\ref{phonon} both for manganites and ferrites with different R ions. As seen, for all compounds the K$_3$ phonon mode is unstable while both $\Gamma^-_2$ and K$_1$ modes are stable. These results indicate that for ferrites systems, similarly as for manganites, the high-temperature P6$_3$/mmc phase transforms directly into the low-temperature P6$_3$cm structure by softening of the K$_3$ zone boundary mode. The non-linear coupling between K$_3$ and the polar $\Gamma^-_2$ mode induces polarization into the system. The polarization is thus a secondary order parameter and the ferroelectricity is improper. Indeed, for a high-temperature structure with frozen-in K$_3$ mode the calculated polarization is very small (less than 0.1 $\mu$C/cm$^2$) and originates only from an electronic contribution. Only when full ionic relaxations are done the polarization of the order of 10 $\mu$C/cm$^2$ develops (see Table.~\ref{phonon}).

\begin{figure}
\begin{center}
\includegraphics[scale=0.3]{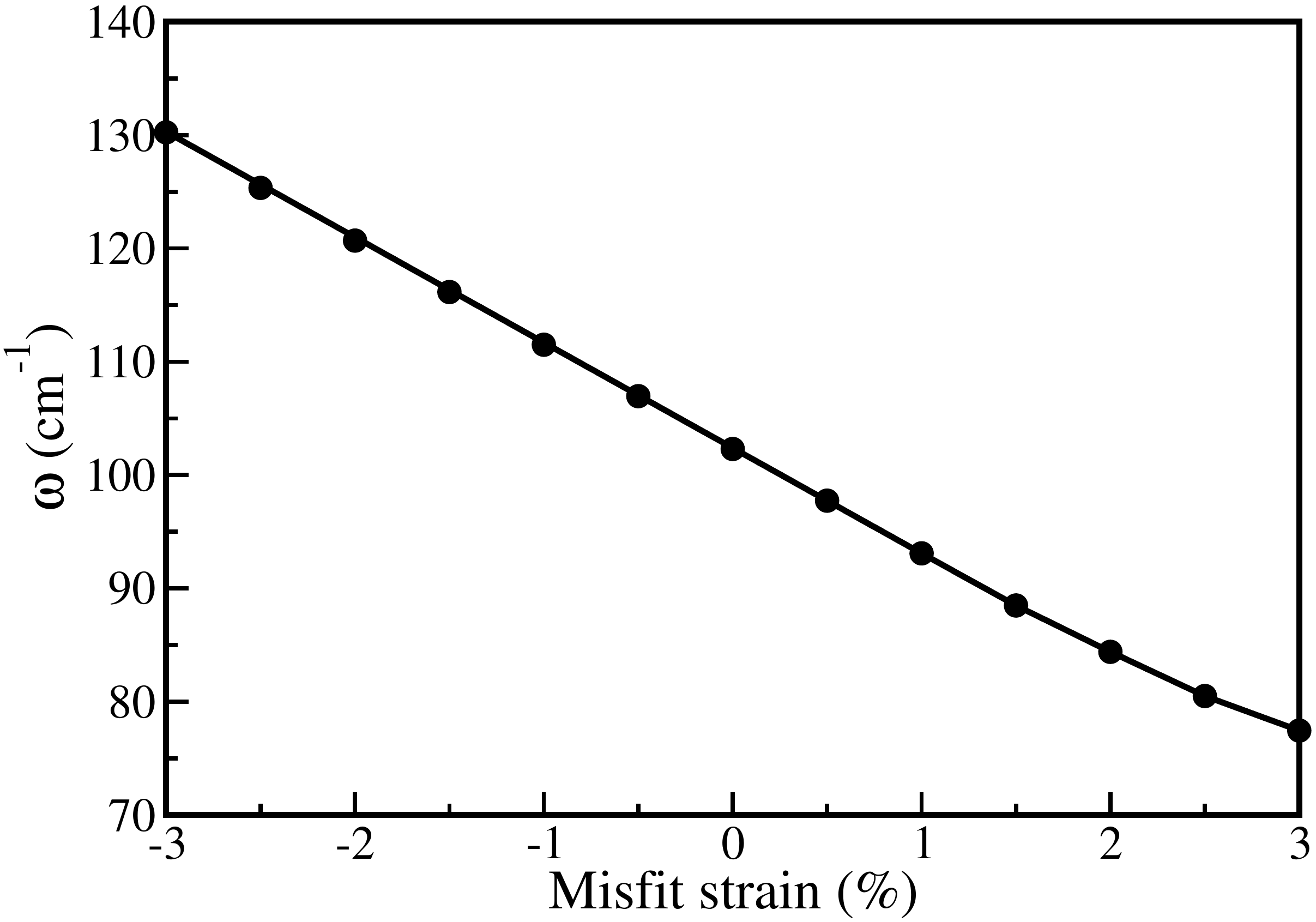}
\end{center}
\caption{\textbf{Effect of epitaxial strain.} Calculated phonon frequency of $\Gamma^-_2$ mode as a function of epitaxial strain.}
\label{strain}
\end{figure}

The above analysis indicates that the origin of ferroelectricity in RFeO$_3$ is the same as in manganites. Therefore, an intermediate P6$_3$mc phase which was proposed to explain the two-step polarization decay observed for YbFeO$_3$ \cite{jeong12b} is rather unlikely. Indeed, the presence of such intermediate phase would typically lead to soft $\Gamma^-_2$ mode which is not the case according to our calculations. In fact, we found that if we freeze-in the $\Gamma^-_2$ mode and perform full structural relaxations, the amplitude of the mode goes to zero. We also found that the $\Gamma^-_2$ mode remains stable for a wide range of epitaxial strain, see Fig. \ref{strain}.

\subsubsection{Non-collinear magnetic spin configurations}
\begin{figure}
\begin{center}
\includegraphics[scale=0.45]{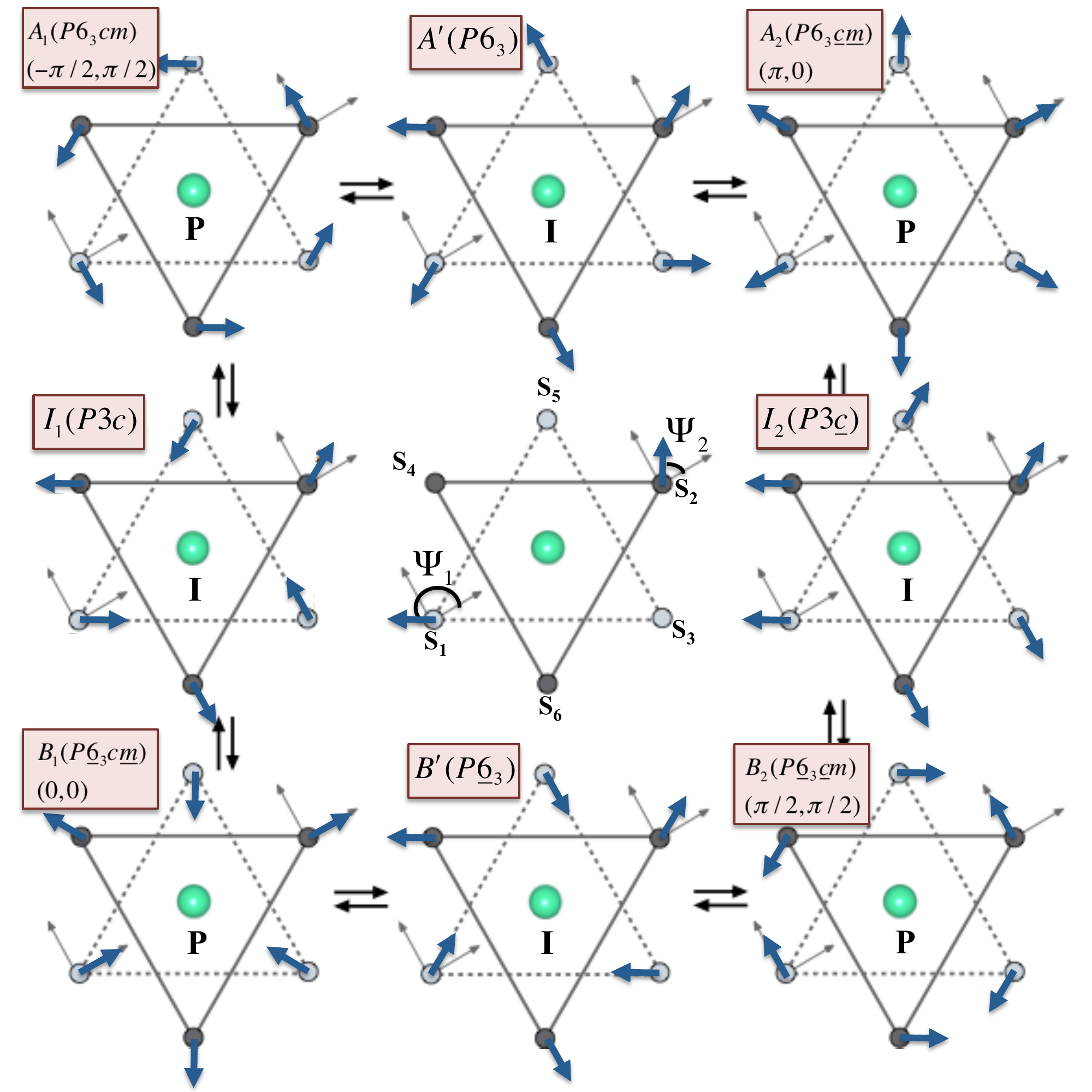}
\end{center}
\caption{\textbf{Non-collinear magnetic structures.} 120$^{\circ}$ spin ordering with spin directions directed by blue arrows. Light and dark gray balls represent magnetic TM ions situated at layer I and II, respectively. The angles $\Psi_1$ and $\Psi_2$ define non-collinear spin configurations. The values of angles $[\Psi_1,\Psi_2]$ in $\Phi = 0$ structural domain are given. Principle magnetic structures, those are compatible with the crystal symmetry, are denoted as A$_1$, A$_2$, B$_1$ and B$_2$. Among four principle magnetic structures A$_2$ allows weak ferromagnetism (wFM). The intermediate states that connect A2 with A1 and B2, denoted as A$^{\prime}$ and I$_2$, respectively, also allow wFM.}
\label{mag-confg}
\end{figure}

We have considered 8 non-collinear magnetic configurations, as depicted in Fig.~\ref{mag-confg}, to determine the magnetic ground state of the hexagonal manganite and ferrite systems.

\subsubsection{Symmetric exchange interaction Determine the magnetic configuration }

\begin{figure*}[t]
\begin{center}
\includegraphics[scale=0.35]{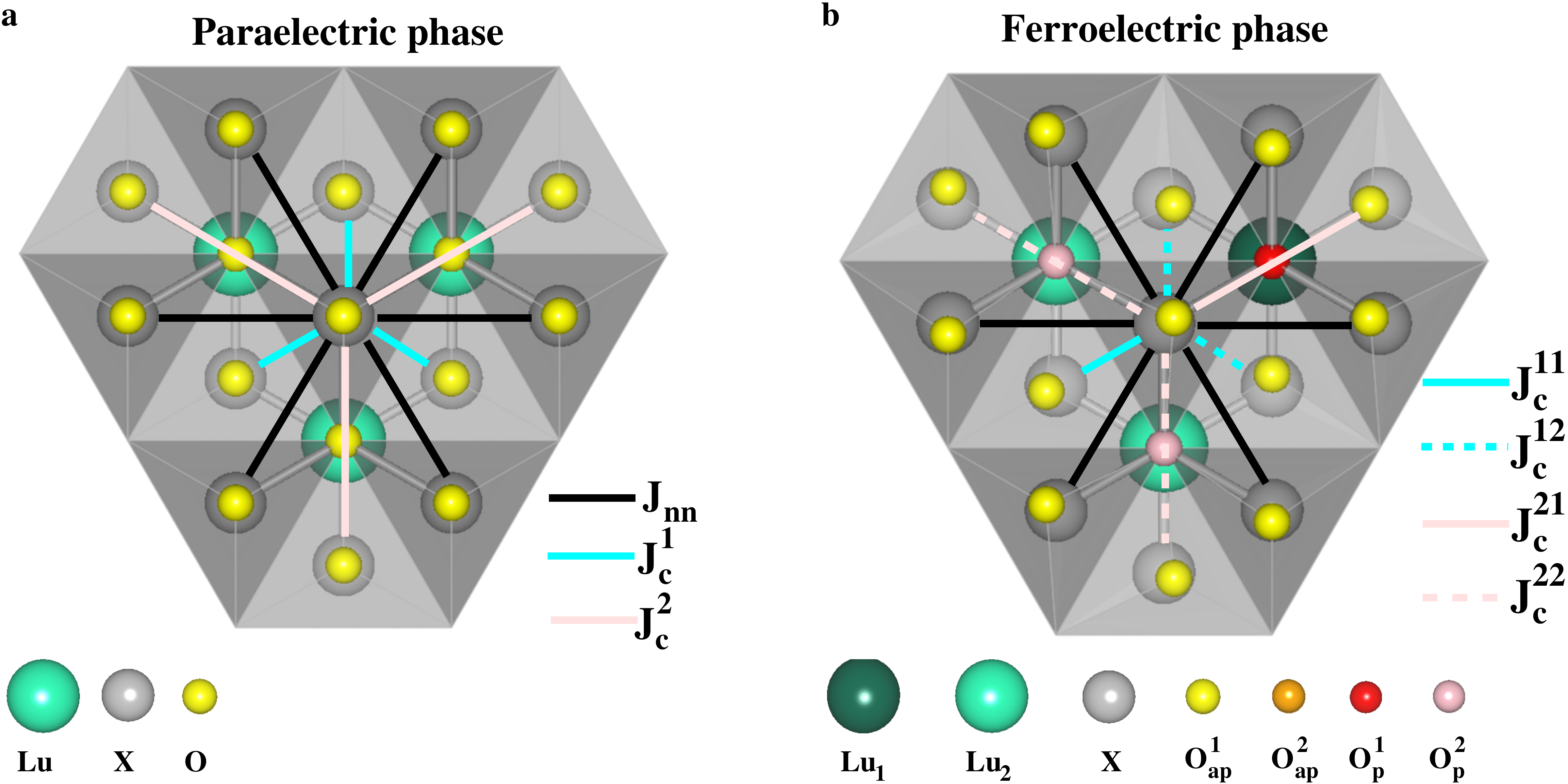}
\end{center}
\caption{\textbf{Symmetric exchange paths} \textbf{a} Symmetric exchange paths in paraelectric phase of $J_{nn}$, $J_{c}^{1}$ and $J_c^2$ are denoted by solid black, cyan and pink lines, respectively. \textbf{b} The trimer distortion splits six fold degenerate nn exchange interaction $J_{nn}$ into two fold degenerate $J_{nn}^1$ (mediated via TM-O$_p^1$-TM exchange path) and four fold degenerate $J_{nn}^2$ (mediated via TM-O$_p^2$-TM exchange path). As the nn exchange interactions are of superexchange in nature and the in-plane distortions in terms of bond lengths and bond angles are not significant, we considered $J_{nn}^1$=$J_{nn}^2$=$J_{nn}$. Similarly trimer distortion splits six fold degenerate inter-layer exchange interactions $J_c^1$ ($J_c^2$) into two fold degenerate $J_c^{11}$ ($J_c^{21}$) and four fold degenerate $J_c^{12}$ ($J_c^{22}$), respectively. While $J_c^{11}$ is mediated by two equivalent X-O-Lu1-O-X exchange pathways, $J_c^{12}$ interaction is mediated by one X-O-Lu1-O-X and one X-O-Lu2-O-X exchange pathways. $J_c^{21}$ and $J_c^{22}$ are mediated via X-O-Lu1-O-X and X-O-Lu2-O-X exchange pathways, respectively.}
\label{J_path}
\end{figure*}

\begin{table}[h]
\caption{Exchange parameters and the corresponding Curie-Weiss temperatures ($\Theta_{CW}$) for LFO and LMO. In the last row (LMO-II) the exchange parameters obtained for LMO using the lattice parameters of LFO are given.}
\begin{center}
\begin{tabular}{lcccccccccc}
\hline
\hline
System&$J_{nn}$&$J_{nnn}$&$J_c^{11}$&$J_c^{12}$&$J_c^{21}$&$J_c^{22}$&$\Theta_{CW}$\\
&meV&meV&meV&meV&meV&meV&K\\
\hline
LFO &6.307 &0.457 &0.942 &0.557 &0.088&0.044&1525\\
\hline
LMO &2.573 &-0.254 &-0.369 &-0.322 &-0.152&0.050&274\\
\hline
LMO-II &3.152 &-0.452 &-0.730 &-0.687 &-0.417&0.113&\\
\hline
\hline
\end{tabular}
\end{center}
\label{J}
\end{table}

We considered in-plane nearest neighbor (nn), $J_{nn}$, and next-nearest neighbor (nnn), $J_{nnn}$, exchange interactions as well as two different inter-layer interactions ($J_c^1$ and $J_c^2$), see Fig. \ref{J_path}. While in the ferroelectric phase $J_{nn}$ splits into two nonequivalent exchange interactions, the difference between them is very small and therefore this effect was neglected. The same applies to $J_{nnn}$. Similarly, $J_c^1$ ($J_c^2$) splits into $J_c^{11}$ and $J_c^{12}$ ($J_c^{21}$ and $J_c^{22}$) interactions. In this case, however, this splitting is essential for determination of the magnetic ground state and therefore it was included. The exchange interactions were calculated by fitting \emph{ab initio} energies of several collinear magnetic configurations. The results are presented in Table \ref{J}.

As seen, $J_{nnn}$ is order of magnitude smaller than $J_{nn}$ and has a negligible influence on magnetic ordering. Therefore, it was neglected in the main manuscript. On the other hand, $J_c^2$ is not always significantly smaller than $J_c^1$. Nevertheless, its qualitative effect is quite analogous to $J_c^1$ and therefore for clarity it was not included in the main text. Here, however, this interaction is included on equal footing with $J_c^1$.

The dominant exchange parameter is $J_{nn}$. It is the AFM superexchange interaction  mediated by equatorial oxygens, O$_{eq}$. For LFO this exchange originates primarily from the hopping from O 2$p$ states into minority Fe 3$d$ states. Since the TM-O$_{eq}$-TM angle is 120$^{\circ}$ the FM superexchange involving two perpendicular $p$ orbitals is expected to be much smaller than the AFM superexchange involving a single orbital. This mechanism is also present for LMO but here the hopping from occupied into empty 3$d$ orbitals is equally important. In the latter case the dominant in-plane hoppings are between $e^{\prime}$ and $e^{\prime\prime}$ orbitals which favor antiparallel spins. As a result $J_{nn}$ for LMO is also AFM. The magnitude of $J_{nn}$ is smaller for LMO. This is partially caused by a larger band gap for this compound as compared to LFO. However, probably more important contribution comes from the hopping from occupied 3$d$ orbitals into empty majority $d_{3z^2}$ states that mediates a FM coupling and partially compensates the AFM contribution. Therefore, ferrites are expected to have in general larger $J_{nn}$ than manganites which indicates that magnetic ordering temperature in ferrites should be much higher; possibly close to the room temperature. 

As seen from Table.~\ref{J}, for LFO the inter-layer exchange parameters are AFM. This can be understood by noting that both exchange parameters originates primarily from virtual hopping of two $p_z$ electrons of O$_{ap}$ ion into 3$d$ minority states of Fe and 5$d$ orbitals of R. These hoppings are the most effective for both apical oxygens in the exchange path only when spins on Fe ions are antiparallel. Otherwise, $p_z$ electrons from different O$_{ap}$ ions cannot hop on the same orbital of the R ion. The above mechanism of AFM interlayer coupling is also present for LMO. In this case however, similarly as for in-plane exchange interactions, there is another type of low-energy excitations that can contribute to the interlayer exchange coupling. They involve hopping from occupied 3$d$ Mn states into empty 3$d$ states of Mn ion in the adjacent plane (note that this hopping is still mediated by O$_{ap}$ and R ions on the exchange path). Clearly, the most important hopping in this case is between filled majority $e^{\prime}$ and empty majority $d_{3z^2}$ which prefers parallel spins. This leads to FM coupling which competes with AFM exchange interaction from the above paragraph. As a result either sign of interlayer exchange couplings is expected for LMO.

Importantly, we find that $|J_c^{k1}| >|J_c^{k2}|$ ($k=1,2$) both for LMO and LFO. In fact this is a generic feature for these systems. Indeed, consider three different nn inter-layer interactions: $J_{12}$, $J_{14}$, and $J_{16}$ (see Fig. \ref{J_path} for notation). In the paraelectric phase they are all equal to $J_c^1$ and their change under the trimer distortion can be written in the following way (up to second order in $Q_{K_3}$):
\begin{widetext}
\begin{equation}
J_{1j}\approx J_c^1+\left(\frac{\partial J_{1j}}{\partial Q_{\rm I}}\right)_{0}Q_{\rm I}+\left(\frac{\partial J_{1j}}{\partial Q^{*}_I}\right)_{0}Q^{*}_1+\left(\frac{\partial^2 J_{1j}}{\partial Q_{\rm I}^2}\right)_{0}Q_{\rm I}^2+\left(\frac{\partial^2 J_{1j}}{\partial (Q_{\rm I}^{*})^2}\right)_{0}(Q_{\rm I}^{*})^2+\left(\frac{\partial^2 J_{1j}}{\partial Q_{\rm I}\partial Q_{\rm I}^{*}}\right)_{0}Q_{\rm I}Q_{\rm I}^{*}
\end{equation}
\end{widetext}
where $j=2,4,6$, subscript $0$ indicates that the derivatives must be evaluated in the paraelectric phase, and order parameters $Q_{\rm I}$, $Q_{\rm I}^*$ were introduced in the previous section. Note that the derivatives must be invariant under the symmetry operations of the paraelectric phase. In particular, this condition requires first order derivatives to be zero and the mixed second order derivatives to be independent on index $j$. Further it imposes relations between remaining second order derivatives such that we can write
\begin{align}
J_{12}\approx J_c^1+cQ_{K_3}^2+2X\cos(2\Phi)Q_{K_3}^2 \\
J_{14}\approx J_c^1+cQ_{K_3}^2+2X\cos(2\Phi+2\pi/3)Q_{K_3}^2 \\
J_{16}\approx J_c^1+cQ_{K_3}^2+2X\cos(2\Phi-2\pi/3)Q_{K_3}^2
\end{align}
where $c$ is a constant and $X=\left(\frac{\partial^2J_{12}}{\partial Q_{\rm I}^2}\right)_{0}$. In the $\alpha$ ($\Phi=0$) phase we have
\begin{align}
J_{12}=J_c^{11}\approx J_c^1+cQ_{K_3}^2+2XQ_{K_3}^2 \\
J_{14}=J_{16}=J_c^{12}\approx J_c^1+cQ_{K_3}^2-XQ_{K_3}^2
\end{align}

It is reasonable to assume that $X$ has the same sign as $J_c^1$ which leads to $|J_c^{11}| >|J_c^{12}|$. Using similar consideration we can also prove that $|J_c^{21}| >|J_c^{22}|$.

 \subsubsection{DM}

\begin{table*}
\caption{Magnitudes of $z$ and $xy$ components of nn DM interactions as well as components of SIA tensor for LFO and LMO compounds. $D^0_z$ is the magnitude of the $z$ component of DM vector in the paraelectric phase while $D_z$ ($D_{xy}$) and $D^{\prime}_z$ ($D^{\prime}_{xy}$) are magnitudes of the $z$ ($xy$) components of DM vectors in the ferroelectric phase mediated by TM-O$^1_p$-TM and TM-O$^2_p$-TM paths, respectively. Notation for components of the SIA tensor is given in the text. Unite are meV.}
\begin{ruledtabular}
\begin{tabular}{ccccccccccccccc}
System &$D_{xy}$ &$D^{\prime}_{xy}$ &$D_z$ &$D^{\prime}_z$ &$D_z^0$ &$\tau_{xx}$ &$\tau_{yy}$ &$\tau_{zz}$ &$\tau_{xz}$ &$\tau^0_{zz}$\\
\hline
LFO  &0.095 &0.097  &0.061 &0.047 &0.072 &0.081   &0.083  &-0.164 &0.018 &0.181 \\
LMO & -0.047  &-0.050 &0.027 &0.033 &0.042 &-0.077 &-0.079 &0.156  &-0.013 &-0.160\\
\end{tabular}
\end{ruledtabular}
\label{parameters2}
\end{table*}

We calculated all six DM interactions acting on TM 1 ion (see Fig.~\ref{DMSIA} for notation). We considered a $\surd 3 \times \surd 3 \times 1$ supercell with 18 RXO$_3$ (X=Mn,Fe) formula units and followed the procedure described in Ref.\cite{Eric12}. This procedure is described as: (1) select one particular interaction by replacing selected X$^{+3}$ ions with Al$^+3$ ions, (2) do total energy calculation by constraining the direction of the spin moments. The $z$ component of the DM interaction between i$^{th}$ and j$^{th}$ TM ions is then given by,

\begin{equation}
D_{ij}^z = \dfrac{1}{S^2}(E[\hat{S}_x^i,\hat{S}_y^j]-E[\hat{S}_{-x}^i,\hat{S}_y^j])
\end{equation}

Similarly one can calculate the other components.

In Table \ref{parameters2} we show the calculated magnitudes of $z$ and $xy$ components of both types of DM vectors for LFO and LMO compounds. The directions of the $z$ components in the paraelectric phase are shown in Fig. \ref{DMSIA}. They are the same for both compounds and they are not altered in the ferroelectric phase. The directions of the transverse components for LFO are shown in Fig. \ref{DMSIA}. For LMO the transverse components are opposite.

 \subsubsection{SIA}

We evaluated the components of SIA tensor by replacing all TM ions by Al$^{+3}$ ions except TM 1 and calculating total energy by constraining spin directions along [100], [010], [001] and [101]. The off-diagonal component $\tau_{xz}$ is given by,

\begin{equation}
\tau_{xz}=\dfrac{1}{2S^2}(2E[\hat{S}_x,0,\hat{S}_z]-E[\hat{S}_x,0,0]-E[0,0,\hat{S}_z])
\end{equation}

The diagonal components were determined by solving the following equations,

\begin{eqnarray}
\tau_{xx}+E_0=\dfrac{1}{S^2}E[\hat{S}_x,0,0]\\
\tau_{yy}+E_0=\dfrac{1}{S^2}E[0,\hat{S}_y,0]\\
\tau_{zz}+E_0=\dfrac{1}{S^2}E[0,0,\hat{S}_z]\\
\tau_{xx}+\tau_{yy}+\tau_{zz}=0
\end{eqnarray}

where $E_0$ is the total energy without any spin-spin interactions.

Calculated components of SIA tensor in Eq. (\ref{SIA}) are shown in Table \ref{parameters2}. The most important  component is $\tau^{zz}$ which is positive for LMO and negative for LFO. Different signs of $\tau^{zz}$ for these materials stems from different number of 3$d$ electrons for these materials. Indeed, in the second order perturbation theory the energy decrease due to SOC can be written as $\Delta E=-\sum_{n\neq m}\lambda^2\frac{|\langle m|\mathbf{\hat s}\cdot\mathbf{\hat l}|n\rangle}{\epsilon_m-\epsilon_n}$. Here $|n\rangle$ and $|m\rangle$ are, respectively, occupied and empty single-electron eigenstates of 3$d$ character with $\epsilon_n$ and $\epsilon_m$ being the corresponding eigenenergies, $\lambda$ is the SOC constant, and $\mathbf{\hat s}$ ($\mathbf{\hat l}$) is single-electron spin (orbital) angular momentum operator. We choose $\mathbf{\hat l}$ quantization axis along $z$. For LFO with half-filled 3$d$ shell we have only spin-flip excitations which are induced by a transverse component of $\mathbf{\hat s}$. In addition, the most important excitations (corresponding to the smallest energy denominators) involve also the change of the magnetic orbital quantum number, $\Delta m_l=\pm1$ which requires a transverse component of $\mathbf{\hat l}$. Therefore, the largest energy gain is achieved when $\mathbf{\hat s}$ and $\mathbf{\hat l}$ are parallel (so that their transverse components couple with each other) preferring spins along $z$ axis and leading to $\tau^{zz}<0$. On the other hand,  for LMO the most important excitation is the transition from majority $e^{\prime}$ to majority $d_{3z^2}$. Here spin is conserved but the magnetic orbital quantum number changes which is possible only when $\mathbf{\hat s}$ and $\mathbf{\hat l}$ are perpendicular to each other, i.e., when spins lie in the $xy$ plane leading to $\tau^{zz}>0$.

~~\\
{\bf Acknowledgments.}
%
We acknowledge discussions with  D.G$.$ Schlom, Julia Mundy, Charles Brooks, P. Schiffer, Maxim Mostovoy, and  J.Moyer. This work started at the suggestion of  D.G$.$S. and J.M. H.D. and C.J.F. were supported by the DOE-BES under Award Number DE-SCOO02334.  A.L.W. was supported by the Cornell Center for Materials Research with funding from NSF  MRSEC program, cooperative agreement DMR-1120296.  Y.G. and W.W. were supported by the U.S. DOE-BES under Award Number DE-SC0008147.\\


\begin{thebibliography}{35}
\expandafter\ifx\csname natexlab\endcsname\relax\def\natexlab#1{#1}\fi
\expandafter\ifx\csname bibnamefont\endcsname\relax
  \def\bibnamefont#1{#1}\fi
\expandafter\ifx\csname bibfnamefont\endcsname\relax
  \def\bibfnamefont#1{#1}\fi
\expandafter\ifx\csname citenamefont\endcsname\relax
  \def\citenamefont#1{#1}\fi
\expandafter\ifx\csname url\endcsname\relax
  \def\url#1{\texttt{#1}}\fi
\expandafter\ifx\csname urlprefix\endcsname\relax\def\urlprefix{URL }\fi
\providecommand{\bibinfo}[2]{#2}
\providecommand{\eprint}[2][]{\url{#2}}

\bibitem[{\citenamefont{Bousquet et~al.}(2008)\citenamefont{Bousquet, Dawber,
  Stucki, Lichtensteiger, Hermet, Gariglio, Triscone, and Ghosez}}]{bousquet08}
\bibinfo{author}{\bibfnamefont{E.}~\bibnamefont{Bousquet}},
  \bibinfo{author}{\bibfnamefont{M.}~\bibnamefont{Dawber}},
  \bibinfo{author}{\bibfnamefont{N.}~\bibnamefont{Stucki}},
  \bibinfo{author}{\bibfnamefont{C.}~\bibnamefont{Lichtensteiger}},
  \bibinfo{author}{\bibfnamefont{P.}~\bibnamefont{Hermet}},
  \bibinfo{author}{\bibfnamefont{S.}~\bibnamefont{Gariglio}},
  \bibinfo{author}{\bibfnamefont{J.-M.} \bibnamefont{Triscone}},
  \bibnamefont{and} \bibinfo{author}{\bibfnamefont{{\relax
  Ph}.}~\bibnamefont{Ghosez}}, \bibinfo{journal}{Nature}
  \textbf{\bibinfo{volume}{452}}, \bibinfo{pages}{732} (\bibinfo{year}{2008}).

\bibitem[{\citenamefont{Tokunaga et~al.}(2009)\citenamefont{Tokunaga, Furukawa,
  Sakai, Taguchi, Arima, and Tokura}}]{tokunaga09}
\bibinfo{author}{\bibfnamefont{Y.}~\bibnamefont{Tokunaga}},
  \bibinfo{author}{\bibfnamefont{N.}~\bibnamefont{Furukawa}},
  \bibinfo{author}{\bibfnamefont{H.}~\bibnamefont{Sakai}},
  \bibinfo{author}{\bibfnamefont{Y.}~\bibnamefont{Taguchi}},
  \bibinfo{author}{\bibfnamefont{T.-h.} \bibnamefont{Arima}}, \bibnamefont{and}
  \bibinfo{author}{\bibfnamefont{Y.}~\bibnamefont{Tokura}},
  \bibinfo{journal}{Nat Mater} \textbf{\bibinfo{volume}{8}},
  \bibinfo{pages}{558} (\bibinfo{year}{2009}).

\bibitem[{\citenamefont{Lee et~al.}(2010)\citenamefont{Lee, Fang, Vlahos, Ke,
  Jung, Kourkoutis, Kim, Ryan, Heeg, Roeckerath et~al.}}]{lee10}
\bibinfo{author}{\bibfnamefont{J.~H.} \bibnamefont{Lee}},
  \bibinfo{author}{\bibfnamefont{L.}~\bibnamefont{Fang}},
  \bibinfo{author}{\bibfnamefont{E.}~\bibnamefont{Vlahos}},
  \bibinfo{author}{\bibfnamefont{X.}~\bibnamefont{Ke}},
  \bibinfo{author}{\bibfnamefont{Y.~W.} \bibnamefont{Jung}},
  \bibinfo{author}{\bibfnamefont{L.~F.} \bibnamefont{Kourkoutis}},
  \bibinfo{author}{\bibfnamefont{J.-W.} \bibnamefont{Kim}},
  \bibinfo{author}{\bibfnamefont{P.~J.} \bibnamefont{Ryan}},
  \bibinfo{author}{\bibfnamefont{T.}~\bibnamefont{Heeg}},
  \bibinfo{author}{\bibfnamefont{M.}~\bibnamefont{Roeckerath}},
  \bibnamefont{et~al.}, \bibinfo{journal}{Nature}
  \textbf{\bibinfo{volume}{466}}, \bibinfo{pages}{954} (\bibinfo{year}{2010}).

\bibitem[{\citenamefont{Tokunaga et~al.}(2012)\citenamefont{Tokunaga, Taguchi,
  Arima, and Tokura}}]{tokunaga12}
\bibinfo{author}{\bibfnamefont{Y.}~\bibnamefont{Tokunaga}},
  \bibinfo{author}{\bibfnamefont{Y.}~\bibnamefont{Taguchi}},
  \bibinfo{author}{\bibfnamefont{T.-h.} \bibnamefont{Arima}}, \bibnamefont{and}
  \bibinfo{author}{\bibfnamefont{Y.}~\bibnamefont{Tokura}},
  \bibinfo{journal}{Nat Phys} \textbf{\bibinfo{volume}{8}},
  \bibinfo{pages}{838} (\bibinfo{year}{2012}).

\bibitem[{\citenamefont{Mermin}(1979)}]{mermin79}
\bibinfo{author}{\bibfnamefont{N.~D.} \bibnamefont{Mermin}},
  \bibinfo{journal}{Rev. Mod. Phys.} \textbf{\bibinfo{volume}{51}},
  \bibinfo{pages}{591} (\bibinfo{year}{1979}).

\bibitem[{\citenamefont{Balke et~al.}(2012)\citenamefont{Balke, Winchester,
  Ren, Chu, Morozovska, Eliseev, Huijben, Vasudevan, Maksymovych, Britson
  et~al.}}]{balke12}
\bibinfo{author}{\bibfnamefont{N.}~\bibnamefont{Balke}},
  \bibinfo{author}{\bibfnamefont{B.}~\bibnamefont{Winchester}},
  \bibinfo{author}{\bibfnamefont{W.}~\bibnamefont{Ren}},
  \bibinfo{author}{\bibfnamefont{Y.~H.} \bibnamefont{Chu}},
  \bibinfo{author}{\bibfnamefont{A.~N.} \bibnamefont{Morozovska}},
  \bibinfo{author}{\bibfnamefont{E.~A.} \bibnamefont{Eliseev}},
  \bibinfo{author}{\bibfnamefont{M.}~\bibnamefont{Huijben}},
  \bibinfo{author}{\bibfnamefont{R.~K.} \bibnamefont{Vasudevan}},
  \bibinfo{author}{\bibfnamefont{P.}~\bibnamefont{Maksymovych}},
  \bibinfo{author}{\bibfnamefont{J.}~\bibnamefont{Britson}},
  \bibnamefont{et~al.}, \bibinfo{journal}{Nat Phys}
  \textbf{\bibinfo{volume}{8}}, \bibinfo{pages}{81} (\bibinfo{year}{2012}).

\bibitem[{\citenamefont{Meier et~al.}(2012)\citenamefont{Meier, Seidel, Cano,
  Delaney, Kumagai, Mostovoy, Spaldin, Ramesh, and Fiebig}}]{meier12}
\bibinfo{author}{\bibfnamefont{D.}~\bibnamefont{Meier}},
  \bibinfo{author}{\bibfnamefont{J.}~\bibnamefont{Seidel}},
  \bibinfo{author}{\bibfnamefont{A.}~\bibnamefont{Cano}},
  \bibinfo{author}{\bibfnamefont{K.}~\bibnamefont{Delaney}},
  \bibinfo{author}{\bibfnamefont{Y.}~\bibnamefont{Kumagai}},
  \bibinfo{author}{\bibfnamefont{M.}~\bibnamefont{Mostovoy}},
  \bibinfo{author}{\bibfnamefont{N.~A.} \bibnamefont{Spaldin}},
  \bibinfo{author}{\bibfnamefont{R.}~\bibnamefont{Ramesh}}, \bibnamefont{and}
  \bibinfo{author}{\bibfnamefont{M.}~\bibnamefont{Fiebig}},
  \bibinfo{journal}{Nature Materials} \textbf{\bibinfo{volume}{11}},
  \bibinfo{pages}{284} (\bibinfo{year}{2012}).

\bibitem[{\citenamefont{Tagantsev and Sonin}(1989)}]{tagantsev89}
\bibinfo{author}{\bibfnamefont{A.~K.} \bibnamefont{Tagantsev}}
  \bibnamefont{and} \bibinfo{author}{\bibfnamefont{E.~B.} \bibnamefont{Sonin}},
  \bibinfo{journal}{Ferroelectrics} \textbf{\bibinfo{volume}{98}},
  \bibinfo{pages}{297} (\bibinfo{year}{1989}).

\bibitem[{\citenamefont{Ramesh and Spaldin}(2007)}]{ramesh07}
\bibinfo{author}{\bibfnamefont{R.}~\bibnamefont{Ramesh}} \bibnamefont{and}
  \bibinfo{author}{\bibfnamefont{N.~A.} \bibnamefont{Spaldin}},
  \bibinfo{journal}{Nat Mater} \textbf{\bibinfo{volume}{6}},
  \bibinfo{pages}{21} (\bibinfo{year}{2007}).

\bibitem[{\citenamefont{Cheong and Mostovoy}(2007)}]{cheong07}
\bibinfo{author}{\bibfnamefont{S.-W.} \bibnamefont{Cheong}} \bibnamefont{and}
  \bibinfo{author}{\bibfnamefont{M.}~\bibnamefont{Mostovoy}},
  \bibinfo{journal}{Nat Mater} \textbf{\bibinfo{volume}{6}},
  \bibinfo{pages}{13} (\bibinfo{year}{2007}).

\bibitem[{\citenamefont{Choi et~al.}(2010)\citenamefont{Choi, Horibe, Yi, Choi,
  Wu, and Cheong}}]{choi10}
\bibinfo{author}{\bibfnamefont{T.}~\bibnamefont{Choi}},
  \bibinfo{author}{\bibfnamefont{Y.}~\bibnamefont{Horibe}},
  \bibinfo{author}{\bibfnamefont{H.~T.} \bibnamefont{Yi}},
  \bibinfo{author}{\bibfnamefont{Y.~J.} \bibnamefont{Choi}},
  \bibinfo{author}{\bibfnamefont{W.}~\bibnamefont{Wu}}, \bibnamefont{and}
  \bibinfo{author}{\bibfnamefont{S.~W.} \bibnamefont{Cheong}},
  \bibinfo{journal}{Nat Mater} \textbf{\bibinfo{volume}{9}},
  \bibinfo{pages}{253} (\bibinfo{year}{2010}).

\bibitem[{\citenamefont{Chae et~al.}(2010)\citenamefont{Chae, Horibe, Jeong,
  Rodan, Lee, and Cheong}}]{chae10}
\bibinfo{author}{\bibfnamefont{S.~C.} \bibnamefont{Chae}},
  \bibinfo{author}{\bibfnamefont{Y.}~\bibnamefont{Horibe}},
  \bibinfo{author}{\bibfnamefont{D.~Y.} \bibnamefont{Jeong}},
  \bibinfo{author}{\bibfnamefont{S.}~\bibnamefont{Rodan}},
  \bibinfo{author}{\bibfnamefont{N.}~\bibnamefont{Lee}}, \bibnamefont{and}
  \bibinfo{author}{\bibfnamefont{S.~W.} \bibnamefont{Cheong}},
  \bibinfo{journal}{Proceedings of the National Academy of Sciences}
  \textbf{\bibinfo{volume}{107}}, \bibinfo{pages}{21366}
  (\bibinfo{year}{2010}).

\bibitem[{\citenamefont{Mostovoy}(2010)}]{mostovoy10}
\bibinfo{author}{\bibfnamefont{M.}~\bibnamefont{Mostovoy}},
  \bibinfo{journal}{Nat Mater} \textbf{\bibinfo{volume}{9}},
  \bibinfo{pages}{188} (\bibinfo{year}{2010}).

\bibitem[{\citenamefont{Kumagai and Spaldin}(2013)}]{kumagai12}
\bibinfo{author}{\bibfnamefont{Y.}~\bibnamefont{Kumagai}} \bibnamefont{and}
  \bibinfo{author}{\bibfnamefont{N.~A.} \bibnamefont{Spaldin}},
  \bibinfo{journal}{Nat Commun} \textbf{\bibinfo{volume}{4}}
  (\bibinfo{year}{2013}).

\bibitem[{\citenamefont{Geng et~al.}(2012)\citenamefont{Geng, Lee, Choi,
  Cheong, and Wu}}]{geng12}
\bibinfo{author}{\bibfnamefont{Y.}~\bibnamefont{Geng}},
  \bibinfo{author}{\bibfnamefont{N.}~\bibnamefont{Lee}},
  \bibinfo{author}{\bibfnamefont{Y.~J.} \bibnamefont{Choi}},
  \bibinfo{author}{\bibfnamefont{S.-W.} \bibnamefont{Cheong}},
  \bibnamefont{and} \bibinfo{author}{\bibfnamefont{W.}~\bibnamefont{Wu}},
  \bibinfo{journal}{Nano Letters} \textbf{\bibinfo{volume}{12}},
  \bibinfo{pages}{6055} (\bibinfo{year}{2012}).

\bibitem[{\citenamefont{{Artyukhin} et~al.}(2012)\citenamefont{{Artyukhin},
  {Delaney}, {Spaldin}, and {Mostovoy}}}]{artyukhin12}
\bibinfo{author}{\bibfnamefont{S.}~\bibnamefont{{Artyukhin}}},
  \bibinfo{author}{\bibfnamefont{K.~T.} \bibnamefont{{Delaney}}},
  \bibinfo{author}{\bibfnamefont{N.~A.} \bibnamefont{{Spaldin}}},
  \bibnamefont{and}
  \bibinfo{author}{\bibfnamefont{M.}~\bibnamefont{{Mostovoy}}},
  \bibinfo{journal}{ArXiv e-prints}  (\bibinfo{year}{2012}),
  \eprint{1204.4126}.

\bibitem[{\citenamefont{Van~Aken et~al.}(2004)\citenamefont{Van~Aken, Palstra,
  Filippetti, and Spaldin}}]{aken04}
\bibinfo{author}{\bibfnamefont{B.~B.} \bibnamefont{Van~Aken}},
  \bibinfo{author}{\bibfnamefont{T.~T.~M.} \bibnamefont{Palstra}},
  \bibinfo{author}{\bibfnamefont{A.}~\bibnamefont{Filippetti}},
  \bibnamefont{and} \bibinfo{author}{\bibfnamefont{N.~A.}
  \bibnamefont{Spaldin}}, \bibinfo{journal}{Nat Mater}
  \textbf{\bibinfo{volume}{3}}, \bibinfo{pages}{164} (\bibinfo{year}{2004}).

\bibitem[{\citenamefont{Fennie and Rabe}(2005)}]{fennie05}
\bibinfo{author}{\bibfnamefont{C.~J.} \bibnamefont{Fennie}} \bibnamefont{and}
  \bibinfo{author}{\bibfnamefont{K.~M.} \bibnamefont{Rabe}},
  \bibinfo{journal}{Phys. Rev. B} \textbf{\bibinfo{volume}{72}},
  \bibinfo{pages}{100103} (\bibinfo{year}{2005}).

\bibitem[{\citenamefont{Bossak et~al.}(2004)\citenamefont{Bossak, Graboy,
  Gorbenko, Kaul, Kartavtseva, Svetchnikov, and Zandbergen}}]{bossak04}
\bibinfo{author}{\bibfnamefont{A.~A.} \bibnamefont{Bossak}},
  \bibinfo{author}{\bibfnamefont{I.~E.} \bibnamefont{Graboy}},
  \bibinfo{author}{\bibfnamefont{O.~Y.} \bibnamefont{Gorbenko}},
  \bibinfo{author}{\bibfnamefont{A.~R.} \bibnamefont{Kaul}},
  \bibinfo{author}{\bibfnamefont{M.~S.} \bibnamefont{Kartavtseva}},
  \bibinfo{author}{\bibfnamefont{V.~L.} \bibnamefont{Svetchnikov}},
  \bibnamefont{and} \bibinfo{author}{\bibfnamefont{H.~W.}
  \bibnamefont{Zandbergen}}, \bibinfo{journal}{Chemistry of Materials}
  \textbf{\bibinfo{volume}{16}}, \bibinfo{pages}{1751} (\bibinfo{year}{2004}).

\bibitem[{\citenamefont{Magome et~al.}(2010)\citenamefont{Magome, Moriyoshi,
  Kuroiwa, Masuno, and Inoue}}]{magome10}
\bibinfo{author}{\bibfnamefont{E.}~\bibnamefont{Magome}},
  \bibinfo{author}{\bibfnamefont{C.}~\bibnamefont{Moriyoshi}},
  \bibinfo{author}{\bibfnamefont{Y.}~\bibnamefont{Kuroiwa}},
  \bibinfo{author}{\bibfnamefont{A.}~\bibnamefont{Masuno}}, \bibnamefont{and}
  \bibinfo{author}{\bibfnamefont{H.}~\bibnamefont{Inoue}},
  \bibinfo{journal}{Japanese Journal of Applied Physics}
  \textbf{\bibinfo{volume}{49}}, \bibinfo{pages}{09ME06}
  (\bibinfo{year}{2010}).

\bibitem[{\citenamefont{Wang et~al.}(2013)\citenamefont{Wang, Zhao, Wang, Gai,
  Balke, Chi, Lee, Tian, Zhu, Cheng et~al.}}]{wang13}
\bibinfo{author}{\bibfnamefont{W.}~\bibnamefont{Wang}},
  \bibinfo{author}{\bibfnamefont{J.}~\bibnamefont{Zhao}},
  \bibinfo{author}{\bibfnamefont{W.}~\bibnamefont{Wang}},
  \bibinfo{author}{\bibfnamefont{Z.}~\bibnamefont{Gai}},
  \bibinfo{author}{\bibfnamefont{N.}~\bibnamefont{Balke}},
  \bibinfo{author}{\bibfnamefont{M.}~\bibnamefont{Chi}},
  \bibinfo{author}{\bibfnamefont{H.~N.} \bibnamefont{Lee}},
  \bibinfo{author}{\bibfnamefont{W.}~\bibnamefont{Tian}},
  \bibinfo{author}{\bibfnamefont{L.}~\bibnamefont{Zhu}},
  \bibinfo{author}{\bibfnamefont{X.}~\bibnamefont{Cheng}},
  \bibnamefont{et~al.}, \bibinfo{journal}{Phys. Rev. Lett.}
  \textbf{\bibinfo{volume}{110}}, \bibinfo{pages}{237601}
  (\bibinfo{year}{2013}).

\bibitem[{\citenamefont{Fiebig et~al.}(2003)\citenamefont{Fiebig, Lottermoser,
  and Pisarev}}]{fiebig03}
\bibinfo{author}{\bibfnamefont{M.}~\bibnamefont{Fiebig}},
  \bibinfo{author}{\bibfnamefont{T.}~\bibnamefont{Lottermoser}},
  \bibnamefont{and} \bibinfo{author}{\bibfnamefont{R.~V.}
  \bibnamefont{Pisarev}}, \bibinfo{journal}{Journal of Applied Physics}
  \textbf{\bibinfo{volume}{93}}, \bibinfo{pages}{8194} (\bibinfo{year}{2003}).

\bibitem[{\citenamefont{Dzyaloshinskii}(1957)}]{dzyaloshinskii57}
\bibinfo{author}{\bibfnamefont{I.~E.} \bibnamefont{Dzyaloshinskii}},
  \bibinfo{journal}{Sov. Phys. JETP} \textbf{\bibinfo{volume}{5}},
  \bibinfo{pages}{1259} (\bibinfo{year}{1957}).

\bibitem[{\citenamefont{Moriya}(1960)}]{moriya60}
\bibinfo{author}{\bibfnamefont{T.}~\bibnamefont{Moriya}},
  \bibinfo{journal}{Phys. Rev.} \textbf{\bibinfo{volume}{120}},
  \bibinfo{pages}{91} (\bibinfo{year}{1960}).

\bibitem[{\citenamefont{Solovyev et~al.}(2012)\citenamefont{Solovyev,
  Valentyuk, and Mazurenko}}]{solovyev12}
\bibinfo{author}{\bibfnamefont{I.~V.} \bibnamefont{Solovyev}},
  \bibinfo{author}{\bibfnamefont{M.~V.} \bibnamefont{Valentyuk}},
  \bibnamefont{and} \bibinfo{author}{\bibfnamefont{V.~V.}
  \bibnamefont{Mazurenko}}, \bibinfo{journal}{Phys. Rev. B}
  \textbf{\bibinfo{volume}{86}}, \bibinfo{pages}{054407}
  (\bibinfo{year}{2012}).

\bibitem[{\citenamefont{Jeong et~al.}(2012{\natexlab{a}})\citenamefont{Jeong,
  Lee, Ahn, and Jang}}]{jeong12a}
\bibinfo{author}{\bibfnamefont{Y.~K.} \bibnamefont{Jeong}},
  \bibinfo{author}{\bibfnamefont{J.-H.} \bibnamefont{Lee}},
  \bibinfo{author}{\bibfnamefont{S.-J.} \bibnamefont{Ahn}}, \bibnamefont{and}
  \bibinfo{author}{\bibfnamefont{H.~M.} \bibnamefont{Jang}},
  \bibinfo{journal}{Chemistry of Materials} \textbf{\bibinfo{volume}{24}},
  \bibinfo{pages}{2426} (\bibinfo{year}{2012}{\natexlab{a}}).

\bibitem[{\citenamefont{Jeong et~al.}(2012{\natexlab{b}})\citenamefont{Jeong,
  Lee, Ahn, Song, Jang, Choi, and Scott}}]{jeong12b}
\bibinfo{author}{\bibfnamefont{Y.~K.} \bibnamefont{Jeong}},
  \bibinfo{author}{\bibfnamefont{J.-H.} \bibnamefont{Lee}},
  \bibinfo{author}{\bibfnamefont{S.-J.} \bibnamefont{Ahn}},
  \bibinfo{author}{\bibfnamefont{S.-W.} \bibnamefont{Song}},
  \bibinfo{author}{\bibfnamefont{H.~M.} \bibnamefont{Jang}},
  \bibinfo{author}{\bibfnamefont{H.}~\bibnamefont{Choi}}, \bibnamefont{and}
  \bibinfo{author}{\bibfnamefont{J.~F.} \bibnamefont{Scott}},
  \bibinfo{journal}{Journal of the American Chemical Society}
  \textbf{\bibinfo{volume}{134}}, \bibinfo{pages}{1450}
  (\bibinfo{year}{2012}{\natexlab{b}}).

\bibitem[{\citenamefont{Akbashev et~al.}(2011)\citenamefont{Akbashev,
  Semisalova, Perov, and Kaul}}]{akbashev11}
\bibinfo{author}{\bibfnamefont{A.~R.} \bibnamefont{Akbashev}},
  \bibinfo{author}{\bibfnamefont{A.~S.} \bibnamefont{Semisalova}},
  \bibinfo{author}{\bibfnamefont{N.~S.} \bibnamefont{Perov}}, \bibnamefont{and}
  \bibinfo{author}{\bibfnamefont{A.~R.} \bibnamefont{Kaul}},
  \bibinfo{journal}{Applied Physics Letters} \textbf{\bibinfo{volume}{99}},
  \bibinfo{eid}{122502} (pages~\bibinfo{numpages}{3}) (\bibinfo{year}{2011}).

\bibitem[{\citenamefont{Fiebig et~al.}(2000)\citenamefont{Fiebig, Fr\"ohlich,
  Kohn, Leute, Lottermoser, Pavlov, and Pisarev}}]{fiebig00}
\bibinfo{author}{\bibfnamefont{M.}~\bibnamefont{Fiebig}},
  \bibinfo{author}{\bibfnamefont{D.}~\bibnamefont{Fr\"ohlich}},
  \bibinfo{author}{\bibfnamefont{K.}~\bibnamefont{Kohn}},
  \bibinfo{author}{\bibfnamefont{S.}~\bibnamefont{Leute}},
  \bibinfo{author}{\bibfnamefont{T.}~\bibnamefont{Lottermoser}},
  \bibinfo{author}{\bibfnamefont{V.~V.} \bibnamefont{Pavlov}},
  \bibnamefont{and} \bibinfo{author}{\bibfnamefont{R.~V.}
  \bibnamefont{Pisarev}}, \bibinfo{journal}{Phys. Rev. Lett.}
  \textbf{\bibinfo{volume}{84}}, \bibinfo{pages}{5620} (\bibinfo{year}{2000}).

\bibitem[{com()}]{comment_U}
\bibinfo{note}{{For LMO $\Theta_{CW}$ is strongly underestimated as compared
  with experimental value of 510 K. This indicates that the value of $U=4.5$ eV
  is too large for LMO. Repeating the calculations for smaller values $U$ we
  reached a reasonable agreement with experiment for $U=2$ eV.}}

\bibitem[{\citenamefont{Anisimov et~al.}(1997)\citenamefont{Anisimov,
  Aryasetiawan, and Lichtenstein}}]{anisimov97}
\bibinfo{author}{\bibfnamefont{V.~I.} \bibnamefont{Anisimov}},
  \bibinfo{author}{\bibfnamefont{F.}~\bibnamefont{Aryasetiawan}},
  \bibnamefont{and} \bibinfo{author}{\bibfnamefont{A.~I.}
  \bibnamefont{Lichtenstein}}, \bibinfo{journal}{Journal of Physics: Condensed
  Matter} \textbf{\bibinfo{volume}{9}}, \bibinfo{pages}{767}
  (\bibinfo{year}{1997}).

\bibitem[{\citenamefont{Perdew et~al.}(1996)\citenamefont{Perdew, Burke, and
  Ernzerhof}}]{PBE}
\bibinfo{author}{\bibfnamefont{J.~P.} \bibnamefont{Perdew}},
  \bibinfo{author}{\bibfnamefont{K.}~\bibnamefont{Burke}}, \bibnamefont{and}
  \bibinfo{author}{\bibfnamefont{M.}~\bibnamefont{Ernzerhof}},
  \bibinfo{journal}{Phys. Rev. Lett.} \textbf{\bibinfo{volume}{77}},
  \bibinfo{pages}{3865} (\bibinfo{year}{1996}).

\bibitem[{\citenamefont{{Kresse, G. and Hafner, J. }}(1993)}]{VASP1}
\bibinfo{author}{\bibnamefont{{Kresse, G. and Hafner, J. }}},
  \bibinfo{journal}{{Phys. Rev. B}} \textbf{\bibinfo{volume}{47}},
  \bibinfo{pages}{558} (\bibinfo{year}{1993}).

\bibitem[{\citenamefont{{Kresse, G. and Furthm\"uller, J. }}(1996)}]{VASP2}
\bibinfo{author}{\bibnamefont{{Kresse, G. and Furthm\"uller, J. }}},
  \bibinfo{journal}{{Phys. Rev. B}} \textbf{\bibinfo{volume}{54}},
  \bibinfo{pages}{11169} (\bibinfo{year}{1996}).

\bibitem[{\citenamefont{Blaha et~al.}()\citenamefont{Blaha, Schwartz, Madsen,
  Kvasnicka, and Luitz}}]{wien2k}
\bibinfo{author}{\bibfnamefont{P.}~\bibnamefont{Blaha}},
  \bibinfo{author}{\bibfnamefont{K.}~\bibnamefont{Schwartz}},
  \bibinfo{author}{\bibfnamefont{G.~K.~H.} \bibnamefont{Madsen}},
  \bibinfo{author}{\bibfnamefont{D.}~\bibnamefont{Kvasnicka}},
  \bibnamefont{and} \bibinfo{author}{\bibfnamefont{J.}~\bibnamefont{Luitz}},
  \bibinfo{note}{wIEN2K, An Augmented Plane Wave + Local Orbitals Program for
  Calculating Crystal Properties, edited by K. Schwarz (Technische
  Universitat{\^a} Wien, Vienna, 2001).}
  
\bibitem{Eric12}C. Weingart, N. Spaldin, E. Bousquet, Phys. Rev. B \textbf{86}, 094413 (2012).

\end{thebibliography}
\end{document}